\documentclass[aps,prd,reprint,onecolumn,10pt,superscriptaddress,nofootinbib,showkeys]{revtex4-2}
\AtBeginDocument{%
}

\usepackage[T1]{fontenc}
\usepackage[utf8]{inputenc}
\usepackage{lmodern}
\usepackage{graphicx}
\usepackage{url}
\usepackage{mathtools}
\usepackage{amssymb}
\usepackage{amsthm}
\usepackage{comment}
\usepackage{xcolor}

\AtBeginDocument{}
\begin{document}

\title{Thermodynamic gravity with non-extensive horizon entropies and topological calibration}

\author{Marco Figliolia}
\email{mfigliolia@unisa.it}
\affiliation{Dipartimento di Fisica `E.R. Caianiello', Universit\`a di Salerno, Fisciano (SA), Italy}

\author{Petr Jizba}
\affiliation{FNSPE, Czech Technical University in Prague, Prague, Czech Republic}

\author{Gaetano Lambiase}
\affiliation{Dipartimento di Fisica `E.R. Caianiello', Universit\`a di Salerno, Fisciano (SA), Italy}
\affiliation{INFN -- Gruppo Collegato di Salerno, Salerno, Italy}

\begin{abstract}
We revisit Jacobson's thermodynamic derivation of gravity for generalized, non-extensive horizon entropies and develop a consistent framework in which modified gravitational dynamics emerge from horizon thermodynamics. By interpreting $S(A)$ as a coarse-grained entropy entering the thermodynamic constitutive relation, we show that only its local slope at a reference area $A_*$ contributes to macroscopic dynamics within a local Rindler-wedge construction. In this setting, the Clausius relation is reformulated as a stationarity condition for a Massieu functional at fixed Unruh temperature, allowing the entropy slope $s_0=(dS/dA)|_{A_*}$ to be identified as the parameter controlling the effective gravitational coupling.
For entropy linear in area, Einstein's equations are recovered with $G_{\rm eff}=1/(4s_0)$. As a separate consistency check, we demonstrate that a scalar-response extension of the horizon entropy density, supplemented by internal entropy production, correctly reproduces the field equations of $f(R)$ gravity.

Motivated by group-entropic arguments and generalized thermodynamics, we adopt a power-law entropy $S(A)=\eta(A/4G)^{\delta}$ and study its implications across scales. To remove the ambiguity in the coarse-graining scale, we introduce a Topological Calibration Principle linking the reference area to intrinsic geometric data through the Gauss--Bonnet theorem. For compact cross-sections admitted by the chosen theory or boundary conditions, this yields a topology-dependent effective coupling governed by the Euler characteristic. Consistency requirements constrain deviations from extensivity and generate scale-dependent modifications with potentially observable cosmological signatures. The resulting framework provides a systematic setting for testing non-extensive horizon thermodynamics and its phenomenological implications.
\end{abstract}

\maketitle

\section{Introduction}
\label{sec:intro}

Over the past two decades, the thermodynamic perspective on gravity has matured into a unifying framework for understanding spacetime dynamics. In this picture, stationary black holes emerge as bona fide thermodynamic systems, which are endowed with a temperature set by the surface gravity and an entropy proportional to the area of the event horizon~\cite{Bek.73,Haw.75,Haw.76,BardeenCarterHawking1973,Smarr1973PRL,Smarr1973PRD}. Central to this viewpoint is the Bekenstein--Hawking relation
\begin{equation}
  S_{\rm BH} \ = \ \frac{A}{4G}\, ,
  \label{eq:SBH}
\end{equation}
where $A$ denotes the horizon area and $G = \ell_{\rm P}^2$ is Newton's gravitational constant in natural units ($\hbar = c = k_{\rm B} = 1$), with $\ell_{\rm P}$ denoting the Planck length. Far from being a mere dimensional estimate, relation~(\ref{eq:SBH}) provides a bridge between classical general relativity, quantum field theory in curved spacetime, and statistical mechanics.

Jacobson in his seminal work~\cite{Jacobson1995} sharpened this connection by demonstrating that the Einstein field equations can be interpreted as an equation of state. Specifically, if one assigns to each local Rindler horizon an entropy density proportional to its area, together with the associated Unruh temperature, then imposing the Clausius relation $\delta Q = T\,dS$ for all local causal horizons leads directly to the Einstein field equations with a cosmological constant~\cite{Jacobson1995,Unruh1976,ParikhSvesko2018}. In this framework, the Einstein equations emerge as macroscopic consistency conditions encoding a local thermodynamic identity of spacetime.

This emergent perspective is conceptually complementary to approaches that treat the metric (or connection or vielbein) as fundamental variables to be quantized. Rather than postulating the graviton as an elementary degree of freedom, one instead seeks a thermodynamic origin of the spacetime metric itself, while remaining agnostic about the precise nature of the underlying constituents --- whether they arise from vacuum entanglement across causal horizons, from collective excitations of a pre-geometric medium, or from more exotic structures~\cite{JacobsonEntanglement2016,Bombelli1986,IorioLambiaseVitiello2004,VanRaamsdonk2010}. 
Closely related developments, particularly those of Padmanabhan and collaborators, emphasize the role of gravitational heat density, the equipartition of degrees of freedom on holographic screens, and the extremization of suitably defined entropy functionals~\cite{Padmanabhan2010,Padmanabhan2005}.
In these frameworks, spacetime dynamics is derived from a variational principle applied to an entropy functional constructed from null vectors, with the Einstein field equations emerging as consistency conditions at its extremum. By contrast, Verlinde's entropic gravity proposal formulates this idea in an explicitly holographic and information-theoretic framework, interpreting Newtonian gravity --- and, in appropriate limits, the Einstein equations --- as entropic forces arising from changes in information on holographic screens endowed with area-scaling entropy~\cite{Verlinde2011}. Despite differences in their technical implementation and in how `emergence' is defined, these approaches share a central conceptual premise: thermodynamic quantities such as entropy, temperature, and heat flux, defined on horizons or screens, are treated as fundamental, while the spacetime metric and its dynamics represent an effective macroscopic response.

On the other hand, semiclassical and/or many-body considerations (further supported by developments in AdS/CFT and AdS/CMT correspondence~\cite{Zaanen2015}) indicate that the simple area law~\eqref{eq:SBH} represents only the leading term in a more general entropic form. In particular, entanglement entropy across a horizon typically exhibits ultraviolet divergences proportional to the area of the entangling surface, but also subleading corrections that depend on field content and background geometry in both field-theoretic and holographic settings~\cite{Srednicki1993,Solodukhin2011,RangamaniTakayanagi2017,CasiniHuertaMyers2011,BuenoMyersPRL2015,BuenoMyersJHEP2015,BlancoCasiniHungMyers2013}.
Moreover, Wald's Noether-charge construction~\cite{WaldGR1984,WaldEntropy} indicates that, in any diffeomorphism-invariant theory of gravity, entropy can be associated with a geometric functional that reduces to the standard area law~\eqref{eq:SBH} in Einstein gravity but acquires additional contributions from higher-curvature terms and topological invariants in more general theories~\cite{WaldEntropy,IyerWald1994}. 
At the same time, systems characterized by long-range interactions and/or strong correlations --- including self-gravitating systems --- often display non-additive thermodynamic behavior and anomalous entropy scaling with system size~\cite{Padmanabhan1990,Chavanis2002,CampaDauxoisRuffo2009}. 
Motivated by these considerations, various non-extensive and/or non-additive generalizations of horizon entropy $S(A)$ have been recently proposed, see, e.g.~\cite{Tsallis1988,TsallisCirto2013,CarusoTsallis2008,JizbaLambiase2022,Moradpour2020,SaridakisBambaMyrzakulovAnagnostopoulos2018,Abbasi2020}.

In gravitational contexts, such generalized entropies are typically introduced at a phenomenological level: the Bekenstein--Hawking entropy $S_{\rm BH}$ is replaced by a modified form $S(A)$, which is then incorporated into an appropriate first-law relation. From this substitution, one infers effective corrections to black hole thermodynamics, the Friedmann equations, or entropic-force--based descriptions of gravity. 
While such models are useful as parametrizations of deviations from general relativity, they leave several conceptual issues open from the perspective of emergent gravity. If gravitational field equations are to arise from horizon thermodynamics, any generalized entropy must be compatible with a local Clausius relation on Rindler patches, as well as with a consistent identification of temperature, energy, and an effective gravitational coupling. In particular, the entropy slope $dS/dA$ plays a central role, as it governs both the macroscopic normalization of horizon entropy and the effective coupling strength in Jacobson-type constructions. From this perspective, it is natural to interpret $S(A)$ as an effective (coarse-grained) constitutive relation for horizon thermodynamics, i.e.\ as a macroscopic input whose local thermodynamic response is fully characterized by the derivative $dS/dA$ evaluated at a reference area $A_*$. In this formulation, the quantity $dS/dA|_{A_*}$ serves as the parameter governing the effective gravitational coupling in the Jacobson-type Clausius relation.

A recent framework proposed by L\"u, Di~Gennaro, and Ong~\cite{LuDiGennaroOng2024} represents an important step in the aforementioned direction. Working within a local-horizon setup in the spirit of Jacobson, the authors introduce a generalized entropy as an effective input within a deformed Legendre structure. This construction leads naturally to a stretched-exponential group entropy functional~\cite{TempestaChaos2020,TsallisJensenPLB2025}, selected so as to ensure the existence of a well-defined canonical temperature on local Rindler patches.
In the spherically symmetric sector, the associated thermodynamic state space reduces effectively to a one-dimensional manifold. As a result, the Clausius relation becomes integrable under suitable regularity assumptions, allowing for a consistent identification of entropy and energy functionals. The resulting equations of state describe an effective Newton coupling that acquires a nontrivial dependence on the horizon area, together with emergent field equations that deviate from the Einsteinian form while preserving a clear thermodynamic interpretation.
This analysis elucidates the extent to which non-extensive entropy functionals can be consistently incorporated within an emergent-gravity framework. At the same time, it highlights that many non-extensive models employed in gravitational physics effectively correspond to one-dimensional truncations, for which the associated integrability constraints are comparatively weak.

In this work, we adopt a complementary and comparatively conservative approach. Instead of undertaking a global analysis of horizon thermodynamics over a multi-parameter 
state space, we restrict our attention to the \emph{local} consequences of a Jacobson-type construction once a non-extensive entropy slope has been specified.  
The only additional input is the choice of a reference area $A_*$ at which the constitutive law is linearized. 
Concretely, we model the entropy associated with a horizon by the phenomenological power-law form $S(A)=\eta(A/4G)^{\delta}$, which appears in the literature under various names, including Tsallis-$\delta$ entropy~\cite{TsallisCirto2013}, Barrow entropy~\cite{Barrow2020}, and Barrow--Tsallis entropy~\cite{Jizba2024}.
In its microcanonical form, this expression is equivalent to the stretched-exponential group entropy~\cite{TempestaChaos2020,TsallisJensenPLB2025}, belonging to the broader class of the so-called group entropies~\cite{Tempesta2016,TempestaGroup2011,TempestaGroup2016,JensenTempestaEntropy2024,JensenJizbaTempestaBH2026}. 
In the following, we will refer to this form simply as \emph{power-law entropy}. 

In the phenomenological part of the analysis we will focus primarily on the sub-area branch $\delta<1$, treating it as a convenient heuristic sector compatible with interpreting the Bekenstein--Hawking area law as a maximal holographic scaling~\cite{Bousso1999,Bousso2002}; the formal construction, however, applies to both sides of $\delta=1$.
 
With the entropic form at hand, we investigate how the associated entropy slope determines the effective Newton coupling on local Rindler wedges. To place our construction on a more systematic footing, we introduce a Topological Calibration Principle (TCP), which relates the reference area at which the entropy slope is evaluated to the intrinsic geometry of horizon cross-sections via the Gauss--Bonnet theorem. In two dimensions, the Gauss--Bonnet theorem furnishes a unique relation between the integrated scalar curvature, the area element, and the Euler characteristic. Employing this relation to fix the calibration scale implements a `no external scales' condition within the emergent-gravity framework and, in settings where non-spherical sectors are admissible, elevates topology to a quantitative criterion constraining admissible deviations from extensivity and/or additivity.

We approach horizon thermodynamics primarily as a framework for ensuring consistency in generalized entropy functionals, rather than as a direct tool for model-building.  At the local level, a canonical Jacobson-type construction identifies the entropy slope as the single macroscopic parameter that determines the effective gravitational coupling in the emergent field equations. This framework recovers Einstein's gravity in the area-law limit. As a separate consistency check, we also show that the same Massieu language is compatible with the standard non-equilibrium scalar-response extension of Jacobson's argument, whose curvature-sensitive realization reproduces field equations of $f(R)$ type.
At the level of calibration, the TCP uses intrinsic curvature and Euler characteristic to calibrate the resolution area, turning $(\eta,\delta)$ into parameters that control the normalization and scale dependence of $G_{\rm eff}$ and, where distinct compact sectors are admissible, its topology dependence.
Taken together, these ingredients lead to explicit logarithmic bounds on $|1-\delta|$ and impose structural constraints on admissible non-extensive horizon entropies.  These bounds become physically meaningful once one assumes a universal pair $(\eta,\delta)$ and identifies the effective coupling inferred from horizon thermodynamics with the low-energy Newton constant. Within this framework, topology provides a quantitative consistency criterion for departures from the Bekenstein--Hawking area law in emergent-gravity scenarios where distinct topological sectors are physically or mathematically admissible.

The paper is organized as follows. In section~\ref{sec:construction} we reformulate Jacobson's local horizon thermodynamics in a canonical form and show that, for area-type entropies, the resulting field equations reduce to the Einstein equations, with $G_{\rm eff}$ directly related to the horizon entropy slope. We also discuss a separate non-equilibrium scalar-response extension, showing how a local response form of the horizon entropy density reproduces the known $f(R)$ result when matched to the Wald entropy density.
In section~\ref{sec:TCP} we introduce the Topological Calibration Principle, which fixes the evaluation scale $A_*$ via the Gauss--Bonnet relation and maps generalized entropy slopes to scale-dependent couplings and, within admissible cross-topology comparisons, topology-dependent couplings $G_{\rm eff}(\chi,A_*)$, leading to explicit bounds on $|1-\delta|$.
Finally, in section~\ref{sec:conclusion} we summarize the main results and discuss their implications for both microscopic models and cosmological phenomenology. For the reader's convenience, we relegate some technical details related to the Gauss--Bonnet entropy to Appendix.

\section{Field equations from local horizon thermodynamics}
\label{sec:construction}

\subsection{Power-law horizon entropy and entropy slope}
\label{subsec:power-law-slope}

A central motivation for exploring generalized horizon entropies is the tension between holographic area laws and standard thermodynamic extensivity. For three-dimensional systems with short--range interactions, ordinary Boltzmann--Gibbs statistical mechanics yields an extensive entropy $S_{\rm BG}\propto V\sim L^{3}$ whenever the number of configurations grows exponentially with the number of constituents, $W(N)\sim \exp(cN)$. Black--hole horizons and many quantum systems with strong correlations instead display an area law, $S\propto A\sim L^{2}$, which is non--extensive in three dimensions~\cite{CarusoTsallis2008,AmicoFazioOsterlohVedral2008}.
In strongly gravitating systems, this discrepancy is accompanied by well-known pathologies of self-gravitating ensembles, including inequivalence between the microcanonical and canonical ensembles, negative heat capacities, and long-lived quasi-stationary states. These features indicate that the Boltzmann--Gibbs framework may become inadequate in such regimes, as indeed it does in practice~\cite{Padmanabhan1990,Chavanis2002,CampaDauxoisRuffo2009}.

In this context, non-additive entropies have been proposed as effective macroscopic descriptions of systems with long-range interactions and/or correlations, including gravitational systems such as black-hole thermodynamics and cosmology~\cite{TsallisCirto2013,CarusoTsallis2008,SaridakisBambaMyrzakulovAnagnostopoulos2018}. 
A particularly economical and structurally well-defined class is provided by Tempesta-type composable group entropies. These entropic functionals preserve the first three Shannon--Khinchin axioms while generalizing the composition law in accordance with the composability principle~\cite{TempestaChaos2020,TsallisJensenPLB2025,TempestaGroup2011,TempestaGroup2016,JensenTempestaEntropy2024}. 
Concretely, one considers functionals of the form
\begin{equation}
  S_{\alpha,\gamma}[p]
  \ = \  \lambda\,\big[S_{R}^{(\alpha)}[p]\big]^{1/\gamma}\, ,
  \label{eq:group-entropy-def}
\end{equation}
where $\lambda$ is a normalization constant, $\gamma>0$ is a deformation parameter, and
\begin{equation}
  S_{R}^{(\alpha)}[p]
  \ \equiv \ \frac{1}{1-\alpha}\,\ln\!\Big(\sum_i p_i^{\alpha}\Big)\, ,
  \label{eq:Renyi-def}
\end{equation}
denotes the Rényi entropy of order $\alpha$.

For a microcanonical ensemble with $W(A)$ equiprobable configurations, labeled by $i=1,\dots,W(A)$, one has $p_i = 1/W(A)$, so that
\begin{eqnarray}
  S_{R}^{(\alpha)}[p_{\rm eq}]
  \ =  \ \frac{1}{1-\alpha}\,
     \ln\!\Bigg(\sum_{i=1}^{W(A)} p_i^{\alpha}\Bigg)
   \ = \  \ln W(A)\, ,
\end{eqnarray}
and therefore
\begin{equation}
  S_{\alpha,\gamma}(A)\ \propto\ [\ln W(A)]^{1/\gamma}\, .
\end{equation}
In this sense, a group-entropic functional adapted to systems with long-range interactions and/or strong correlations, when evaluated on a microcanonical horizon ensemble, generically leads to power-law deformations of the Bekenstein--Hawking area law. To make this more precise, suppose that the number of microstates associated with a horizon area exhibits a sub-exponential scaling of the form~\cite{HanelThurnerEPL2011}
\begin{equation}
  W(A)\ \sim\ \exp\!\Big[c\,\Big(\frac{A}{4G}\Big)^{q}\Big]\, ,
  \qquad c>0\, ,\quad q>0 \, ,
\end{equation}
so that
\begin{equation}
  \ln W(A)\ \propto\ \Big(\frac{A}{4G}\Big)^{q}\, .
\end{equation}
Evaluating the group-entropy functional~\eqref{eq:group-entropy-def} on this microcanonical ensemble then yields
\begin{equation}
  S_{\alpha,\gamma}(A)\ \propto \ \Big[\ln W(A)\Big]^{1/\gamma}
  \ \propto\ \Big(\frac{A}{4G}\Big)^{q/\gamma}\, .
\end{equation}
In this framework the ratio $q/\gamma$ plays the role of an effective non-extensivity index, while $\lambda$ and numerical factors can be absorbed into an overall normalization.

Guided by these considerations, and without committing to a specific underlying microscopic model, we adopt the phenomenological power-law constitutive relation
\begin{equation}
  S(A) \ = \ \eta\,\Big(\frac{A}{4G}\Big)^{\delta},
  \qquad \eta \ > \ 0,\quad \delta \ > \ 0\, ,
  \label{eq:entropy-law}
\end{equation}
with $\eta$ and $\delta$ playing the role of effective entropic indices that encode possible departures from the Bekenstein--Hawking law. In the units $c=\hbar=k_B=1$ used throughout, the combination $A/4G$ is dimensionless, so that $S$ is dimensionless and both $\eta$ and $\delta$ are pure numbers. This makes~\eqref{eq:entropy-law} a dimensionally consistent deformation of the Bekenstein--Hawking relation~\eqref{eq:SBH}. The Bekenstein--Hawking case corresponds to $\eta=1$ and $\delta=1$, while $\delta\neq1$ describes non--extensive deformations of the area law in the sense discussed above. In a schematic microcanonical picture one may formally identify $\delta = q/\gamma$ and absorb proportionality constants into $\eta$, but in what follows we treat $(\eta,\delta)$ as phenomenological parameters to be constrained by consistency conditions rather than fixed by a specific microstate counting.

For concreteness, in our phenomenological discussion we focus primarily on the regime
\[
  \frac{1}{2} \ < \ \delta \ \leq \ 1\, ,
\]
which corresponds to sub-area deviations from the Bekenstein--Hawking law. This choice is not motivated by a microscopic necessity, but rather defines a convenient heuristic branch. 
For large horizon areas, the entropy grows more slowly than the Bekenstein--Hawking area law, thereby treating the latter as a maximal holographic scaling rather than as a relation to be exceeded~\cite{Bousso1999,Bousso2002}.
Equivalently, for $0<\delta<1$ one has
\[
  \frac{d^2S}{dA^2} \ < \ 0\, ,
\]
so that the entropy slope decreases with the coarse-graining area, as expected for an effective description in which long-range correlations or holographic redundancies reduce the number of independent horizon degrees of freedom.

The canonical construction itself requires only that $S(A)$ be differentiable and monotonic; hence, the formalism applies for any $\delta>0$. The opposite branch $\delta>1$, which includes some relevant generalized-entropy cosmology models (e.g. Barrow model~\cite{Barrow2020} or Tsallis--Cirto model~\cite{TsallisCirto2013}), is therefore not excluded in our formalism. It simply corresponds to super-area scaling and is more strongly constrained if the Bekenstein--Hawking law is interpreted as a holographic upper bound. Accordingly, our constraints will be expressed in terms of $|1-\delta|$, thereby covering both branches $\delta<1$ and $\delta>1$.


With this phenomenological horizon law in place, we now turn to its local consequences. The central question is how a given entropy slope
\begin{equation}
  s_\delta(A_*)\ \equiv \ \frac{dS}{dA}\Big|_{A_*}\, ,
\end{equation}
defined at a reference (coarse-graining) area scale $A_*$, translates into an effective gravitational coupling on local Rindler patches. In the local construction, the near-horizon expansion is controlled by patch data (e.g. the affine cutoff $\lambda_c$ and the local area element $dA$), while $A_*$ enters only through the macroscopic parameter $s_\delta(A_*)$.

In particular, in the sub-area branch used for most of our phenomenological discussion here, we have

\begin{equation}
  S(A) \ = \ \eta\Big(\frac{A}{4G}\Big)^{\delta}\, ,\qquad \eta \ > \ 0\, ,\quad \tfrac12 \ < \ \delta \ \le \ 1\, ,
  \label{eq:entropy-law-local}
\end{equation}
so that
\begin{equation}
  s_\delta(A_*) \ = \ \eta\,\delta\,(4G)^{-\delta}\,A_*^{\delta-1}\, ,
\end{equation}
and the canonical construction below will reproduce
\begin{equation}
  G_{\rm eff}(A_*) \ = \ \frac{1}{4\,s_\delta(A_*)}
  \ = \ \frac{G}{\eta\,\delta}\left(\frac{4G}{A_*}\right)^{\delta-1}\, ,
\end{equation}
in agreement with the running-$G$ picture of \cite{LuDiGennaroOng2024}.

\subsection{Local Rindler wedge and optical expansion}
\label{subsec:rindler-optics}

We work on a smooth, time-oriented Lorentzian manifold $(\mathcal M,g_{ab})$ with Levi-Civita connection and curvature conventions
\begin{eqnarray}
  &&R^a{}_{bcd} \ = \  \partial_c \Gamma^a_{bd} \ - \  \partial_d \Gamma^a_{bc}
  \ + \ \Gamma^a_{ce}\,\Gamma^e_{bd} \ - \  \Gamma^a_{de}\,\Gamma^e_{bc}\, ,\nonumber \\[2mm]
  &&R_{bd} \ = \  R^a{}_{bad}\, ,\qquad
  R \ = \ g^{bd}R_{bd}\, .
  \label{eq:Riemann-conv-local}
\end{eqnarray}
Metric variations are taken with respect to the contravariant components $g^{ab}$, and we use the shorthand
\begin{equation}
  \delta_g F \ \equiv \ \left.\frac{d}{d\epsilon}F[g(\epsilon)]\right|_{\epsilon=0}\, ,
  \qquad
  \delta g^{ab} \ \equiv \ \left.\frac{d}{d\epsilon}g^{ab}(\epsilon)\right|_{\epsilon=0}\, ,
\end{equation}
for a one-parameter family $g_{ab}(\epsilon)$ with $g_{ab}(0)=g_{ab}$.

To implement the local construction, we first recall the geometry of a local Rindler wedge. Around any spacetime point $p$ one can construct a wedge whose boundary $\mathcal H$ is generated by an affinely parametrized null congruence $k^a$ with affine parameter $\lambda$ such that $\lambda=0$ at $\Sigma$. In Jacobson's setup we consider the portion of $\mathcal H$ to the past of $\Sigma$, hence $\lambda\in[-\lambda_c,0]$ with $\lambda_c>0$. 
We choose a spacelike two-surface $\Sigma$ through $p$ as a local bifurcation surface, and pick a wedge-preserving boost Killing field $\chi^a$ that vanishes on $\Sigma$ and is tangent and null on $\mathcal H$. 
To be more specific, one can, for instance, consider local inertial coordinates $(T,X,Y,Z)$ centered at $p$.  One may then take $\Sigma$ to be a small spacelike two-disk at $T=X=0$, parametrized by the transverse coordinates $(Y,Z)$. The associated local causal horizon is then generated, to leading order at $p$, by the null vector $k^a=(\partial_T+\partial_X)^a$, with $\lambda$ chosen as an affine parameter along these generators and normalized so that $\lambda=0$ on $\Sigma$.

\begin{figure}[t]
  \centering
  \includegraphics[width=0.80\textwidth]{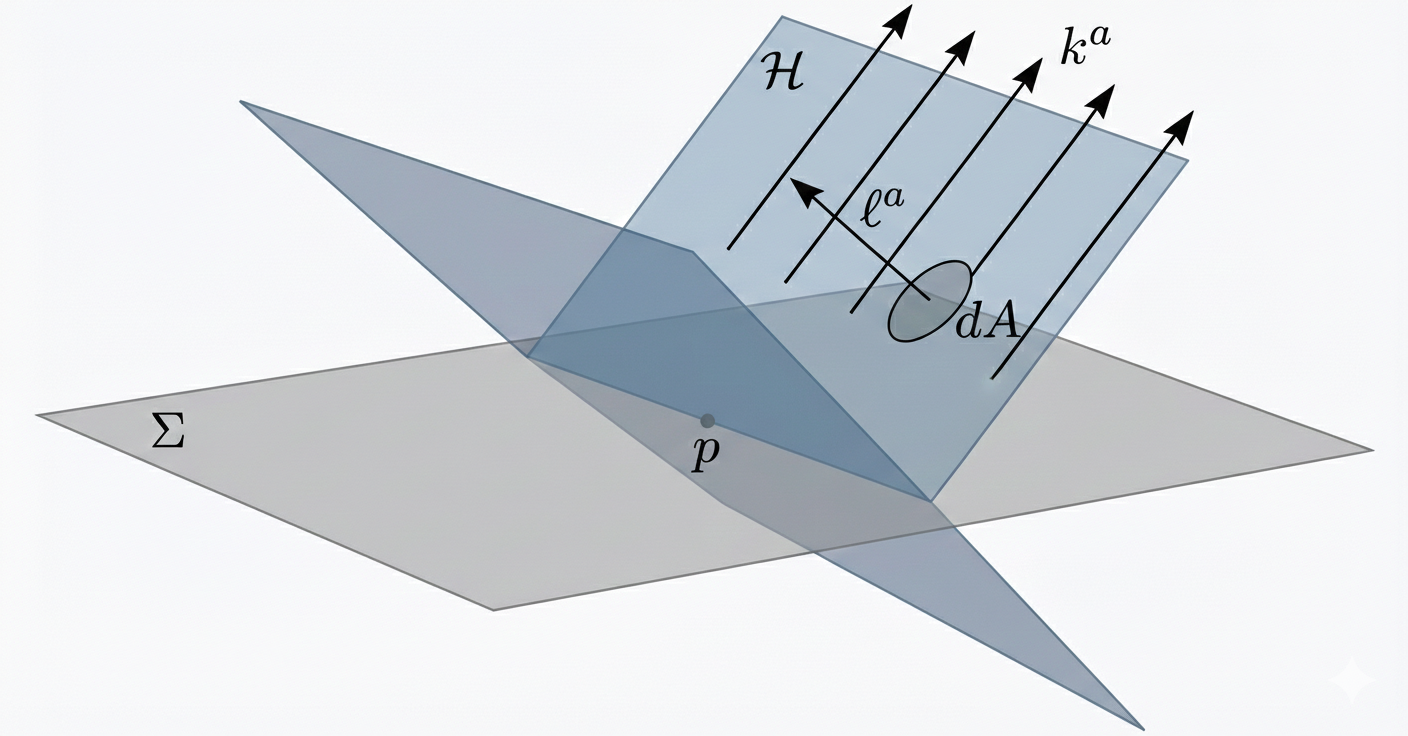}
 \caption{ Local Rindler-wedge geometry near a spacetime point $p$ (with $p\in\Sigma$). The boundary $\mathcal{H}$ is a null hypersurface generated by an affinely parametrized null congruence $k^a$, emanating from the spacelike local cross-section $\Sigma$. An auxiliary null vector $\ell^a$ fixes the transverse direction, and $dA$ denotes the induced area element on the spacelike two-dimensional cuts of $\mathcal{H}$.}
  \label{fig:rindler-wedge}
\end{figure}

With this normalization, near $\Sigma$ one has
\begin{equation}
  \chi^a \ = \  -\,\kappa\,\lambda\,k^a \ + \  \mathcal{O}(\lambda^2),
  \qquad
  T \ = \  \frac{\kappa}{2\pi}\, ,\quad \beta \ = \ \frac{2\pi}{\kappa}\, ,
  \label{eq:chi-k-local}
\end{equation}
where $T$ is the local Unruh temperature associated with the boost flow and $\beta$ the corresponding inverse temperature~\cite{Unruh1976}. 
The hypersurface element on $\mathcal H$ is
\begin{equation}
  d\Sigma^b \ = \  k^b\ d\lambda\ dA\, ,
\end{equation}
with $dA$ the induced area element on the spacelike two-dimensional cross-sections of $\mathcal H$ at fixed affine parameter $\lambda$. The resulting local Rindler-wedge geometry and notation are depicted in figure~\ref{fig:rindler-wedge}.

To describe the focusing of the null generators, we introduce an auxiliary null vector $\ell^a$ with $k^a\ell_a=-1$ and the projector onto the screen space
\begin{equation}
  h_{ab} \ = \ g_{ab} \ + \ k_a\ell_b \ + \ \ell_a k_b\, .
\end{equation}
The optical tensor
\begin{equation}
  B_{ab} \ \equiv \ h_a{}^{c}h_b{}^{d}\nabla_c k_d\, ,
\end{equation}
decomposes into expansion, shear and twist as
\begin{equation}
  B_{ab} \ = \ \tfrac12\theta\,h_{ab} \ + \ \sigma_{ab} \ + \ \omega_{ab}\, ,
\end{equation}
where $\theta \equiv B^a{}_{a}$, $\sigma_{ab}$ is the trace-free symmetric part and $\omega_{ab}$ the antisymmetric part. 
Since $k^a$ is chosen to generate the null hypersurface $\mathcal H$ (hence it is hypersurface-orthogonal), Frobenius'' theorem implies the integrability condition $k_{[a}\nabla_b k_{c]}=0$, and therefore the twist (vorticity) tensor vanishes, $\omega_{ab}=0$ (see e.g.\ \cite{WaldGR1984,PoissonToolkit2004}).

For an affine congruence $k^b\nabla_b k^a=0$ the Raychaudhuri equation acquires the form 
\begin{equation}
  \frac{d\theta}{d\lambda}
  \ =\  -\,\tfrac12\ \!\theta^2 \ -\  \sigma_{ab}\sigma^{ab}
        \ -\  R_{ab}\,k^a k^b\, ,
  \label{eq:Raychaudhuri-local}
\end{equation}
while the shear obeys the Sachs equation, projected on the screen space
\begin{equation}
  \frac{d\sigma_{ab}}{d\lambda}
  \ = \  -\ \!\theta\,\sigma_{ab}
    \ - \  \mathcal{C}_{ab}
    \ - \ \Big(\sigma_{ac}\sigma^{c}{}_{b}
           \ - \  \tfrac{1}{2}h_{ab}\,\sigma_{cd}\sigma^{cd}\Big)\, .
  \label{eq:Sachs-local}
\end{equation}
Here $\mathcal{C}_{ab}$ denotes the screen-projected (trace-free) tidal term along $k^a$, expressed in terms of the Weyl tensor $C_{abcd}$ as
\begin{equation}
  \mathcal{C}_{ab}
  \ \equiv \ h_a{}^{c} h_b{}^{d} C_{ecfd}\,k^e k^f\, .
\end{equation}
By construction, $\mathcal{C}_{ab}$ is symmetric and trace--free with respect to $h_{ab}$, so the right--hand side of~\eqref{eq:Sachs-local} is itself trace--free and thus compatible with the definition of $\sigma_{ab}$ as the trace--free part of the optical tensor $B_{ab}$.

The expansion in the affine parameter is a local-equilibrium expansion around the local cross-section $\Sigma$, where the approximate boost Killing field vanishes. The cutoff $\lambda_c$ is chosen small compared with the local curvature radius and with the length scale over which the matter flux varies appreciably along the generators. The leading non-trivial contribution to the integrated area change is then of order $\lambda_c^2$ and is controlled by the Ricci focusing term.
Terms beyond this order are finite-size corrections to the local patch, or corrections associated with departures from local equilibrium, and we neglect them in the equilibrium Einstein branch. Equivalently, the field equation is extracted from the universal leading term in the limit of an arbitrarily small local Rindler wedge.

If we now impose the initial conditions~\cite{Jacobson1995}
\begin{equation}
  \theta|_\Sigma \ = \ 0\, , \qquad \sigma_{ab}|_\Sigma \ = \  0\, ,
\end{equation}
and assume that the congruence is affinely parametrized and that \(R_{ab}\) and the optical data \((\theta,\sigma_{ab},\omega_{ab})\) are smooth in a neighborhood of \(p\), then equation~\eqref{eq:Sachs-local} implies that \(\sigma_{ab} = O(|\lambda|)\) near \(\lambda = 0\), and hence \(\sigma_{ab}\sigma^{ab} = O(\lambda^2)\). Substituting this into equation~\eqref{eq:Raychaudhuri-local} shows that
\begin{equation}
  \theta^2 \;\;\; \mbox{and} \;\;\; \sigma_{ab}\sigma^{ab} \ = \ \mathcal{O}(\lambda^2)\, ,
\end{equation}
and hence, to first non-trivial order in $\lambda$,
\begin{equation}
  \frac{d\theta}{d\lambda} \ = \  -\,R_{ab}k^a k^b \ + \ \mathcal{O}(|\lambda|)\, .
\end{equation}
Expanding $\theta(\lambda)$ near $\Sigma$ for sufficiently small affine cutoff $\lambda_c$ (i.e. for $|\lambda|\le\lambda_c$)
\begin{equation}
  \theta(\lambda) \ = \ \theta|_\Sigma \ + \ \lambda\Big.\frac{d\theta}{d\lambda}\Big|_\Sigma \ + \ \mathcal{O}(\lambda^2)\, ,
\end{equation}
and using $\theta|_\Sigma=0$ gives
\begin{equation}
  \theta(\lambda) \ = \ -\,\lambda\,R_{ab}k^a k^b \ + \ \mathcal{O}(\lambda^2)\, .
  \label{eq:theta-linear}
\end{equation}
Since the area element on the cuts changes along the generators according to the standard definition of the expansion
\begin{equation}
  \theta \ \equiv \  \frac{1}{dA}\,\frac{d(dA)}{d\lambda}
  \quad \Rightarrow \quad
  \frac{d(dA)}{d\lambda} \ = \  \theta\ dA\, ,
\end{equation}
equation~\eqref{eq:theta-linear} implies, to first non-trivial order in $\lambda$
\begin{equation}
  \frac{d(dA)}{d\lambda}
  \ = \  -\,\lambda\,R_{ab}k^a k^b\,dA \ + \  \mathcal{O}(\lambda^2)\, .
  \label{eq:dA-variation}
\end{equation}
Integrating along the generators over the past segment $\lambda\in[-\lambda_c,0]$ then yields the linear area response of the horizon patch
\begin{equation}
\delta_g\!\Big(\!\int_{\mathcal{H}} dA\Big)
  \ = \  -\!\int_{-\lambda_c}^{0}\!\lambda\,R_{ab}k^a k^b\,d\lambda\,dA
  \ +\ \mathcal{O}(\lambda_c^3)\, ,
  \label{eq:area-patch-local}
\end{equation}
so the leading variation of the area element along the generators is controlled entirely by the Ricci focusing term $R_{ab}k^a k^b$, and the corresponding integrated area change over $\lambda\in[-\lambda_c,0]$ scales as $\mathcal{O}(\lambda_c^2)$. The quadratic optical terms $\theta^2$ and $\sigma_{ab}\sigma^{ab}$ enter Raychaudhuri at $\mathcal{O}(\lambda^2)$ and therefore correct $\theta(\lambda)$ only at $\mathcal{O}(\lambda^3)$, yielding contributions to the integrated area change starting at $\mathcal{O}(\lambda_c^4)$. These terms are beyond the leading local-equilibrium order used to derive the Einstein branch.

\subsection{Canonical Massieu balance and the Einstein branch}
\label{subsec:massieu-einstein}

In this setup, the boost field $\chi^a$ also plays the role of modular generator~\cite{CasiniTesteTorroba2017}. The Bisognano--Wichmann theorem implies that, in Minkowski spacetime, the vacuum state reduced to a Rindler wedge is thermal at inverse temperature $\beta=2\pi/\kappa$ with respect to the boost charge~\cite{BisognanoWichmann1975,BisognanoWichmann1976}
\begin{equation}
  K \ = \ \int_{\mathcal H}\chi^a T_{ab}\,d\Sigma^b\, .
\end{equation}
Using the near-horizon expansion~\eqref{eq:chi-k-local} and $d\Sigma^b=k^b\,d\lambda\,dA$, one finds, to $\mathcal{O}(\lambda_c^2)$
\begin{equation}
K
  \ = \  -\,\kappa\!\int_{-\lambda_c}^{0}\!\lambda\,T_{ab}k^a k^b\,d\lambda\,dA
    \ + \  \mathcal{O}(\lambda_c^3)\, ,
  \label{eq:K-local-again}
\end{equation}
so that, within the local Rindler approximation, the matter state restricted to the wedge is represented by a density operator thermal with respect to the boost generator,
$\rho\propto e^{-\beta K}$ with $\beta=2\pi/\kappa$.

Here $\rho$ denotes the matter state restricted to the local Rindler wedge or, in the usual density-operator language, the reduced density operator associated with the wedge degrees of freedom. It should not be interpreted as a classical mass density, energy density, or perfect-fluid variable. The stress tensor $T_{ab}$ is therefore kept arbitrary; when a semiclassical interpretation is needed, it may be understood as the expectation value of the matter stress tensor in the wedge state $\rho$. No perfect-fluid decomposition of the form $T_{ab}=(\varrho+p)u_a u_b+p g_{ab}$ is assumed. The local balance uses only the null projection $T_{ab}k^a k^b$, i.e. the boost-energy flux through the horizon patch.

To phrase the local dynamics in canonical form it is convenient to introduce the Massieu (also known as Legendre) functional
\begin{equation}
  \Psi[g;\beta] \ =\  S_G[g] \ -\  \beta\,\langle K\rangle_\rho\, ,
  \label{eq:Massieu}
\end{equation}
where $S_G[g]$ is the gravitational (horizon) entropy functional and $\langle K\rangle_\rho$ denotes the expectation value of $K$ in the fixed wedge state $\rho$. In the present canonical setting we vary only the geometry at fixed inverse temperature and fixed matter state
\begin{equation}
  \delta\beta \ = \ 0\, ,\qquad \delta\rho\ = \ 0\,,
\end{equation}

The condition $\delta\rho=0$ indicates that the quantum state associated with the wedge is held fixed when comparing nearby patch geometries. In particular, this condition does not impose a fluid equation of state, nor does it restrict the algebraic structure of $T_{ab}$.

We then require stationarity with respect to patch-localized metric variations:
\begin{equation}
  \delta_g\Psi \ = \ 0 \qquad\Leftrightarrow \qquad
  \delta_g S_G \ = \ \beta\,\delta_g\langle K\rangle_\rho\,  .
  \label{eq:stationarity}
\end{equation}

Throughout the variation $\delta_g$ we keep fixed the thermodynamic scheme: the inverse temperature $\beta$, the matter state $\rho$, and the reference scale $A_*$ entering $s_0=dS/dA|_{A_*}$. In particular, $\delta_g$ acts on the patch geometry, not on the calibration point $A_*$. This implements local canonical equilibrium under $\delta_g\beta=\delta_g\rho=0$. We also choose the variation to preserve, to leading order, the affine parametrization of the generators $k^a$, the normalization $k\!\cdot\!\ell=-1$, and the initial local-equilibrium data $\theta|_{\Sigma}=\sigma_{ab}|_{\Sigma}=0$. In particular, we require that the approximate boost Killing field in the varied geometry $g_{ab}(\epsilon)$ be normalized in the same way as in equation~\eqref{eq:chi-k-local}, namely
\begin{equation}
  \chi^a(\epsilon) \ = \ -\,\kappa\,\lambda\,k^a(\epsilon) \ + \ \mathcal{O}(\lambda^2)\, ,
  \label{eq:chi-k-eps}
\end{equation}
with the same surface gravity $\kappa$.

The fixed thermodynamic data $(\beta,\rho,A_*)$, together with the leading-order conditions on the local horizon generators stated above and the boost normalization~\eqref{eq:chi-k-eps}, specify the family of nearby local Rindler patches over which the stationarity condition~\eqref{eq:stationarity} is evaluated. The corresponding metric variations should not be understood as unrestricted variations of an action principle, nor as a statement that the full spacetime causal structure is held fixed under arbitrary variations. Rather, these assumptions keep fixed only the auxiliary local Rindler-frame structure needed to compare nearby horizon patches within the same thermodynamic scheme.

Within this restricted family, the metric variation remains physically non-trivial. At the order retained in the small-$\lambda$ expansion, it probes the Ricci focusing term $R_{ab}k^a k^b$ for an otherwise arbitrary null direction $k^a$ at the point $p$. Requiring the resulting null-projected identity to hold for every null vector and every spacetime point is sufficient to reconstruct the tensorial field equation up to the standard trace, or integration-constant, ambiguity.

Differentiating at $\epsilon=0$ one gets
\begin{equation}
  \delta\chi^a
  \ \equiv \ \left.\frac{d\chi^a(\epsilon)}{d\epsilon}\right|_{\epsilon=0}
  \ = \  \mathcal{O}(\lambda^2)\, ,
\end{equation}
so that the additional contribution of $\delta\chi^a$ to $\delta_g\langle K\rangle$ scales as $\int_{-\lambda_c}^{0}\lambda^2\,d\lambda = \mathcal{O}(\lambda_c^3)$ and is sub-leading with respect to the $\mathcal{O}(\lambda_c^2)$ terms retained in the Raychaudhuri expansion.

In what follows, we use the shorthand notation
\begin{equation}
  \delta_g\langle K\rangle
  \ \equiv \  \left.\frac{d}{d\epsilon}\,\langle K[g(\epsilon)]\rangle\right|_{\epsilon=0}\, ,
\end{equation}
for variations of the boost energy with respect to the geometry at fixed matter state $\rho$.

In the area-type branch, we assume that, to leading order on the patch, the horizon entropy density (per unit transverse area) is constant,
\begin{equation}
  s(x) \ \equiv \ s_0\, .
  \label{eq:area-type-density}
\end{equation}

We stress that $s(x)$ here refers to the entropy of the horizon itself, not to a matter or fluid density. The constant $s_0$ is not an additional independent parameter: it is simply the local notation for the entropy slope introduced above. More explicitly,

\begin{equation}
  s_0 \ \equiv \ s_\delta(A_*) \ \equiv \ \frac{dS}{dA}\Big|_{A_*}\, ,
  \label{eq:s0-def}
\end{equation}
where $A_*$ is the coarse-graining area scale at which the global entropy law is linearized.
In this case
\begin{equation}
  S_G[g] \ = \ \int_{\mathcal H}s_0\,dA\, ,\qquad
  \delta_g S_G \ = \ s_0\,\delta\!\!\int_{\mathcal H}\!dA\, ,
\end{equation}
and equation~\eqref{eq:area-patch-local} gives
\begin{equation}
  \delta_g S_G
  \ = \  -\,s_0\!\int_{-\lambda_c}^{0}\!\lambda\,R_{ab}k^a k^b\,d\lambda\,dA
    \ + \ \mathcal{O}(\lambda_c^3)\, .
  \label{eq:deltaS-local}
\end{equation}
On the `heat' side, at fixed \(\beta\), the variation of the boost charge yields the heat flux across the patch. Using~\eqref{eq:K-local-again} together with the expansion~\eqref{eq:chi-k-eps}, and adopting the same \(\mathcal{O}(\lambda_c)\) power counting as above, we evaluate \(\delta_g\langle K\rangle\) at fixed matter state \(\rho\). In this case, metric-induced variations of \(T_{ab}\) contribute only at orders beyond the \(\mathcal{O}(\lambda_c^2)\) accuracy retained here\footnote{Operationally, the condition $\delta\rho=0$ implies that the local excitation content of the wedge is held fixed while the geometry of the patch is varied. Any induced response of the stress tensor $T_{ab}$ then contributes to $\beta \,\delta_g\langle K\rangle$ only at higher order in the small-$\lambda$ expansion and therefore does not enter the leading-order null-projected balance.}. One then finds
\begin{equation}
  \beta\,\delta_g\langle K\rangle
  \ = \ -\,2\pi\!\int_{-\lambda_c}^{0}\!\lambda\,T_{ab}k^a k^b\,d\lambda\,dA
    \ + \ \mathcal{O}(\lambda_c^3)\, .
  \label{eq:deltaK-local}
\end{equation}
With the convention $\lambda\in[-\lambda_c,0]$ and $k^a$ future-directed, the integrand $-\lambda\,T_{ab}k^a k^b$ is non-negative when $T_{ab}k^a k^b\ge0$, so positive energy flux corresponds to positive heat into the wedge. equation~\eqref{eq:deltaK-local} is the boost-energy flux across the patch at fixed $\beta$.

Putting together the entropy and heat variations, and inserting
\eqref{eq:deltaS-local} and \eqref{eq:deltaK-local} into the
stationarity condition~\eqref{eq:stationarity}, one finds
\begin{equation}
-\,s_0\!\int_{-\lambda_c}^{0}\!\lambda\,R_{ab}k^a k^b\,d\lambda\,dA
  \ = \ -\,2\pi\!\int_{-\lambda_c}^{0}\!\lambda\,T_{ab}k^a k^b\,d\lambda\,dA
  \ +\ \mathcal{O}(\lambda_c^3)\, .
\end{equation}

The common factor \(\int_{-\lambda_c}^{0}\lambda\,d\lambda = -\lambda_c^2/2\) is nonzero (and identical on both sides) and therefore cancels, while the \(\mathcal{O}(\lambda_c^3)\) corrections are neglected when working to \(\mathcal{O}(\lambda_c^2)\) in the Raychaudhuri expansion. One is thus left with the pointwise null identity.
\begin{equation}
  s_0\,R_{ab}k^a k^b \ = \  2\pi\,T_{ab}k^a k^b
  \;\;\;\;\text{for every null }\;\;\;\;k^a\text{ at }p\, ,
  \label{eq:null-eq}
\end{equation}
which is the key local statement: stationarity of the Massieu functional ties the null-projected Ricci focusing to the null energy flux, and it holds for every null direction \(k^{a}\) at \(p\).
The overall sign follows from the choice of orientation of the horizon generators and from the convention that positive \(T_{ab}k^{a}k^{b}\) corresponds to energy flux into the wedge. A simultaneous reversal \(k^{a}\!\to\!-k^{a}\), together with the corresponding affine reparametrization, leaves~\eqref{eq:null-eq} invariant, so the ensuing tensor equation is independent of this convention.

To make the resulting tensor structure explicit, it is convenient to define
\begin{equation}
  H_{ab} \ \equiv \ s_0\,G_{ab} \ - \ 2\pi\,T_{ab}\, ,
\end{equation}
and to recall that, for null vectors, $G_{ab}k^a k^b=R_{ab}k^a k^b$. Equation~\eqref{eq:null-eq} then implies $H_{ab}k^a k^b=0$ for all null $k^a$.
A standard algebraic lemma in Lorentzian geometry (see e.g.~\cite{WaldGR1984}) states that any symmetric tensor with vanishing double contraction on all null vectors must be proportional to the metric\footnote{A short proof is obtained by working in a local orthonormal frame and using linear combinations of null vectors to show that all eigenvalues of $H^a{}_b$ coincide, so that $H_{ab}=\Lambda_{\rm loc}\,g_{ab}$ for some scalar $\Lambda_{\rm loc}$.}
\begin{equation}
  H_{ab} \ = \ \Lambda_{\rm loc}\,g_{ab}\, , 
\end{equation}
for some scalar function $\Lambda_{\rm loc}$. 
Taking the covariant divergence and using $\nabla^aG_{ab}=0=\nabla^aT_{ab}$ gives $\nabla_b\Lambda_{\rm loc}=0$, hence $\Lambda_{\rm loc}=\Lambda$ is a constant.

We stress that the condition $\nabla^aT_{ab}=0$ is an additional macroscopic assumption, not a consequence of the null-projected algebraic relation alone. It selects the divergence-free, or continuity-preserving, sector considered in the present work: the local boost-energy flux is described by the conventional covariantly conserved stress tensor, and the light-cone statistical measure is kept fixed.

This is the sense in which the present derivation corresponds to a particular macroscopic/statistical completion of the null-projected dynamics. In the light-cone statistical framework of \cite{IsidroPaganiniPesci2025LightCone}, unimodular gravity is interpreted as a canonical-type ensemble, while modified-measure theories correspond to a grand-canonical-type extension in which the relevant measure, or density of gravitational states, is allowed to vary. Our construction remains in the divergence-free, fixed-measure sector analogous to the standard Jacobson/unimodular completion, and does not introduce additional measure degrees of freedom.

If the condition $\nabla^aT_{ab}=0$ is relaxed, the local thermodynamic argument still fixes the null-projected content of the field equations, equivalently the traceless part up to the usual scalar ambiguity, but the trace sector and the effective cosmological term are not determined by the same argument alone. They require additional statistical data specifying the non-continuity-preserving completion. Such generalized modified-measure or grand-canonical-type sectors are beyond the scope of the present paper.

Dividing by $s_0$ one finally arrives at the Einstein field equations
\begin{equation}
  G_{ab} \ = \ 8\pi\,G_{\rm eff}\,T_{ab} \ + \ \Lambda g_{ab}\, ,
  \qquad 
  G_{\rm eff} \ = \ \frac{1}{4\,s_0}\, .
  \label{eq:Einstein-unified}
\end{equation}
In the area-type branch, the canonical stationarity condition for
\(\Psi = S_G - \beta \langle K \rangle\) on local Rindler patches reduces to Einstein's equations, with the Newton coupling fixed by the leading entropy response on the patch. When the horizon entropy density is constant at this order, \(S_G = \int_{\mathcal H} s_0\, dA\), the gravitational input to the thermodynamic balance is fully captured by the single parameter
\(s_0=s_\delta(A_*)=\left.dS/dA\right|_{A_*}\). 
The corresponding effective coupling then follows as
\(G_{\rm eff} = 1/(4 s_0)\).

\subsection{Non-equilibrium scalar-response extension}
\label{subsec:scalar-response-extension}

It is useful at this point to separate the minimal area-type branch used in the TCP construction from possible non-equilibrium extensions. In the derivation above $s(x)$ was taken to be constant along the generators at leading order, so that only the local slope $s_0=s_\delta(A_*)$ entered the equilibrium Massieu balance. This is the only ingredient needed for the topology- and scale-dependent bounds derived below.

If instead one allows the horizon entropy density to be governed by a local, non-constant scalar response, the equilibrium Clausius relation is no longer sufficient. A convenient way to parametrize this possibility, without inserting curvature at the outset, is to write

\begin{equation}
  s(x) \ = \ \eta\,\Phi(x)\, ,\qquad \eta>0\, ,
  \label{eq:s-scalar-response}
\end{equation}

where $\Phi(x)$ represents an effective macroscopic scalar response variable on the horizon patch. We do not construct a separate dynamical scalar-tensor model for $\Phi$ in this paper; rather, $\Phi$ parametrizes a possible non-equilibrium response of the horizon entropy density. When $\Phi(x)$ varies along the generators, $k^a \nabla_a s \neq 0$ already at $\lambda=0$, and the local Clausius relation must be supplemented by an internal entropy-production term. This scalar-response form should be understood as a separate non-equilibrium extension of the local horizon construction, not as an additional ingredient in the TCP analysis.

The familiar $f(R)$ field equations emerge only if one assumes that this phenomenological response matches the Wald entropy density of a local, curvature-based effective action. In such a specific case, one sets

\begin{equation}
  \Phi(x) \ = \ f'\!\big(R(x)\big)\, ,
  \qquad
  s(x) \ = \ \eta\,f'\!\big(R(x)\big)\, ,
  \label{eq:s-curv-short}
\end{equation}

a choice directly motivated by Wald's Noether-charge formula for $f(R)$ theories, in which the horizon entropy density is proportional to $f'(R)$~\cite{WaldEntropy,IyerWald1994}. Thus curvature is not part of the minimal entropy law used in the TCP analysis; it enters only in this standard scalar-response realization, after matching to the Wald density of a curvature-based effective action.\footnote{See \cite{WaldEntropy,IyerWald1994} for the derivation from the action functional; here we only use that, for $f(R)$ theories, the Wald entropy density associated with a horizon cross-section is proportional to $f'(R)$.}

With this choice one has generically $k\!\cdot\!\nabla s=\eta\,k\!\cdot\!\nabla f'(R)\neq0$ already at $\lambda=0$, so the local Clausius relation
\begin{eqnarray}
  \delta Q \ = \  T\,\delta S\, ,
\end{eqnarray}
is no longer integrable on the patch: the entropy change associated with a given geometric perturbation depends on the microscopic path followed along the generators. Following the non-equilibrium analysis of Eling, Guedens and Jacobson~\cite{ElingGuedensJacobson2006}, one introduces an internal production term $d_i S$ and works instead with
\begin{equation}
  \delta Q \ = \  T\,\big(\delta S \ + \  d_i S\big)\, ,
  \label{eq:NE-balance-short}
\end{equation}
where $d_i S$ is local on the patch, collects the $\mathcal{O}(\lambda^2)$ contributions that are quadratic in the departures from equilibrium, and is non-negative in the near-equilibrium regime.

In what follows we only sketch the steps needed to extract the null-projected identity~\eqref{eq:null-fR-short}, referring to \cite{ElingGuedensJacobson2006} for the full non-equilibrium derivation.
In a small-$\lambda$ expansion, and imposing again $\theta|_{\Sigma}=\sigma_{ab}|_{\Sigma}=0$, an appropriate choice of $d_i S$ (including terms of the form $\sigma^2$, $\theta^2$ and $(k\!\cdot\!\nabla f')^2$) allows one to isolate the linear contributions in $\lambda$ that are sensitive to $\delta g^{ab}$. Quadratic pieces are reinterpreted as entropy production and absorbed into $d_i S$.

On the heat side, the calculation is identical to the area-type case: under the same assumptions on $\delta\chi^a$ and on the initial data, the variation of the boost charge at fixed $\beta$ still yields equation~\eqref{eq:deltaK-local}, with the same $\mathcal{O}(\lambda_c^3)$ remainder.

On the entropy side, specializing the scalar response to the Wald-matched realization $s(x)=\eta f'(R)$ and expanding at small $\lambda$, one finds that the linear terms in $\lambda$ can be expressed, after an integration by parts along the generators and the use of Raychaudhuri at $\Sigma$, as a combination of
\begin{eqnarray}
  f'(R)\,R_{ab}k^a k^b
  \quad\text{and}\quad
  k^a k^b \nabla_a\nabla_b f'(R)\, ,
\end{eqnarray}
multiplied by $\eta$.
Inserting these contributions into the non--equilibrium balance~\eqref{eq:NE-balance-short} and simplifying the common nonzero factor $\int_{-\lambda_c}^{0}\lambda\,d\lambda$, one obtains, using the arbitrariness of $k^a$ at $p$, the null-projected identity
\begin{equation}
  \frac{2\pi}{\eta}\,T_{ab}k^a k^b
  \ =\  \big[f'(R)R_{ab} \ - \ \nabla_a\nabla_b f'(R)\big]\,k^a k^b\, .
  \label{eq:null-fR-short}
\end{equation}
Exactly as in the area-type branch, one completes the tensor by defining
\begin{eqnarray}
\mathcal E_{ab} \ \equiv \ f'(R)R_{ab} \ - \ \nabla_a\nabla_b f'(R) \ - \ \frac{2\pi}{\eta}\,T_{ab}\, ,
\end{eqnarray}
which satisfies $\mathcal E_{ab}k^a k^b=0$ for every null $k^a$. The algebraic lemma discussed below equation~\eqref{eq:null-eq} then implies $\mathcal E_{ab}=\Upsilon\,g_{ab}$ for some scalar function $\Upsilon$. 
Taking the covariant divergence, using the Bianchi identities and $\nabla^aT_{ab}=0$, and noting that $\nabla_b f'(R)=f''(R)\nabla_b R$, one uniquely fixes
$\Upsilon=\Box f'(R)-\tfrac12 f(R)+\Lambda$, so that the full field equations read
\begin{equation}
  f'(R)R_{ab} \ - \ \nabla_a\nabla_b f'(R)
  \ + \ \big(\Box f'(R) \ - \ \tfrac12 f(R)\big)g_{ab}
  \ = \ \frac{2\pi}{\eta}\,T_{ab}\, .
  \label{eq:fR-final-short}
\end{equation}
Identifying $2\pi/\eta = 8\pi G_{\rm eff}$, equation~\eqref{eq:fR-final-short} coincides with the Euler--Lagrange equations obtained by varying the action
\begin{equation}
\label{eq:action-f(R)}
    I \ = \ \frac{1}{16\pi G_{\rm eff}}\!\int\! d^4x\,\sqrt{-g}\,f(R) \ + \ I_{\rm m}[g,\psi]\, ,
\end{equation}
where $I_{\rm m}[g,\psi]$ is the matter action. In particular, choosing $G_{\rm eff}=G$ and $\eta=1/(4G)$ reproduces the Wald entropy density for $f(R)$ gravity~\cite{WaldEntropy,IyerWald1994}. 

This subsection therefore has a limited and precise role: it shows that the canonical Massieu framework is compatible with Jacobson's known non-equilibrium extension. The area-type horizon entropy density used in the TCP analysis leads directly to Einstein's equations with $G_{\rm eff}=1/(4s_0)$, whereas the Wald-matched curvature-sensitive realization $s(x)\propto f'(R)$, supplemented by the appropriate internal production term $d_i S$, reproduces the familiar $f(R)$ field equations in agreement with \cite{ElingGuedensJacobson2006,BrusteinHadad2009,SotiriouFaraoni2010,DeFeliceTsujikawa2010}. No curvature-dependent horizon entropy density is used in deriving the TCP bounds below.

In the following, we return to the area-type branch~\eqref{eq:entropy-law-local}, which is the branch used in the TCP construction, and treat $s_0$ as the fundamental parameter controlling the local effective coupling via equation~\eqref{eq:Einstein-unified}, namely $G_{\rm eff}=1/(4s_0)$.
The remaining task is to calibrate the resolution scale $A_*$, and hence $s_0$, in a non--arbitrary way, using only intrinsic data on the horizon cross section. In the next section we formulate this requirement as a Topological Calibration Principle, based on the Gauss--Bonnet relation between area, curvature and Euler characteristic, and use it to translate generalized entropy slopes into topology- and scale-dependent constraints on admissible non-extensive exponents.

\section{Topological Calibration Principle}
\label{sec:TCP}
\subsection{Minimal calibration and reference area}
\label{subsec:TCP-minimal}

We now examine how intrinsic topology can calibrate the reference scale entering the local Jacobson construction. Throughout this section, $\Sigma$ denotes a compact, boundaryless horizon cross section (or a compact quotient thereof) introduced solely for calibration; it should not be conflated with the local spacelike two-surface used to construct the Rindler wedge in section~\ref{sec:construction}. 
As shown in the previous section, once the entropy slope $s_0$ defined in equation~\eqref{eq:s0-def} is specified, horizon thermodynamics fixes the effective coupling in equation~\eqref{eq:Einstein-unified} as $G_{\rm eff}=1/(4s_0)$. It is therefore natural to regard the reference area scale $A_*$ appearing in $s_0$ as fixed by intrinsic data of the chosen cross section, rather than as an external input.

The use of topology in this construction is deliberately restricted to the horizon sectors admitted by the theory or boundary conditions. We do not assume that arbitrary compact topologies are generically realized as stationary astrophysical black-hole horizons in four-dimensional asymptotically flat general relativity. Under standard Hawking-type topology theorems and related rigidity or topology-censorship results, the allowed horizon topology is strongly constrained, with spherical cross sections providing the relevant black-hole case. Non-spherical sectors become physically admissible only when changing the global setting, the boundary conditions, or the gravitational dynamics, for example in AdS topological black holes, compact quotients, or modified-gravity models.
An explicit example of such a modified-gravity setting is provided by \cite{NampalliwarSuvorovKokkotas2020}, where topologically non-trivial black-hole solutions in $f(R)$ gravity are constructed and their possible electromagnetic signatures are discussed.

Accordingly, the TCP should be understood as a calibration scheme and consistency test within such admitted classes of horizon cross sections. If the underlying theory or boundary conditions admit a given family of compact sectors, and if their effective couplings are compared at a fixed area-averaged intrinsic curvature scale $|\tilde R_*|$, then the Gauss--Bonnet theorem fixes the corresponding reference area entering $s_0$. The resulting topology dependence of $G_{\mathrm{eff}}$ is therefore a quantitative constraint produced by applying the same entropy-slope prescription across admitted sectors, rather than a claim about generic horizon topology in ordinary four-dimensional general relativity.

This motivates a \emph{minimal calibration principle} for $A_*$, according to which the reference area must (i) be constructed from intrinsic invariants of the cross section, (ii) introduce no macroscopic scale beyond those present in the induced geometry, and (iii) remain stable under smooth deformations within a given topological class. In this section, we assume that the entropy parameters $(\eta,\delta)$ appearing in the power-law entropy~\eqref{eq:entropy-law-local} are universal, i.e., independent of both the horizon topology and the scale at which the slope $s_0$ in equation~\eqref{eq:s0-def} is evaluated; whenever this assumption is relaxed (for instance by allowing $\eta(\chi)$) we will state it explicitly. 

For definiteness, and consistently with the conservative sub-area branch discussed in section~\ref{sec:construction}, we display the TCP bounds mainly in the window $\tfrac{1}{2}<\delta\le 1$.

Combining the entropy law~\eqref{eq:entropy-law-local} with the definition of $s_0$ then gives
\begin{equation}
  s_0 \ \equiv \ s_\delta(A_*) 
      \ = \  \eta\,\delta\,(4G)^{-\delta} A_*^{\delta-1}\, ,
  \label{eq:slope-delta}
\end{equation}
so that the effective Newton constant entering the local Einstein equations becomes
\begin{equation}
  G_{\rm eff}(A_*) 
  \ = \  \frac{1}{4 s_\delta(A_*)}
  \ = \  \frac{G}{\eta \delta}\left(\frac{4G}{A_*}\right)^{\delta-1}\, .
  \label{eq:Geff-Astar}
\end{equation}
For $\delta=1$ and $\eta=1$ this reproduces the standard Bekenstein--Hawking values $s_1=1/(4G)$ and $G_{\rm eff}=G$. For $\delta\neq1$ the same non-extensive scaling implies a scale dependence of the thermodynamic coupling, with $A_*$ serving as the reference (coarse-graining) area at which the entropy slope is evaluated. Formally, this is the same structural mechanism emphasized in \cite{LuDiGennaroOng2024}: generalized horizon entropies determine an effective Newton constant proportional to the inverse entropy slope, $G_{\rm eff}\propto s_0^{-1}$.

To make the geometric aspects of the calibration explicit, consider a smooth, orientable, spacelike two-surface $\Sigma$ representing a horizon cross section, with induced metric $\gamma_{ij}$, area element $\sqrt{\gamma}\,d^2x$, and total area
\begin{equation}
  A \ = \  \int_{\Sigma} \!\sqrt{\gamma}\  d^2x \,.
\end{equation}
The Gauss--Bonnet theorem then reads
\begin{equation}
  \int_{\Sigma}\!\sqrt{\gamma}\ \ {}^{(2)}\!R\ d^2x \ = \  4\pi\,\chi(\Sigma)\, ,
  \label{eq:GB-2D-TCP}
\end{equation}
where $\chi(\Sigma)$ is the Euler characteristic of $\Sigma$ (for an orientable surface of genus $g$, $\chi(\Sigma) = 2-2g$), and ${}^{(2)}\!R$ denotes the intrinsic scalar curvature of $(\Sigma, \gamma_{ij})$.

It is convenient to encode the intrinsic scale of a given cross-section $\Sigma$ in the area-averaged scalar curvature
\begin{equation}
  \tilde R_{\Sigma} \ \equiv \ \frac{1}{A}\int_{\Sigma}\!\sqrt{\gamma}\ \ {}^{(2)}\!R\ d^2x\, ,
  \label{eq:Rbar-def}
\end{equation}
which equals twice the mean Gaussian curvature. In two dimensions this quantity plays a distinguished role: it is a natural global scalar with dimensions of inverse length squared, linear in ${}^{(2)}\!R$ and directly tied to the Gauss--Bonnet invariant. 
Combining equation~\eqref{eq:GB-2D-TCP} with the definition~\eqref{eq:Rbar-def} gives
\begin{equation}
  \tilde R_{\Sigma}\ = \ \frac{4\pi\,\chi(\Sigma)}{A}\,,
  \label{eq:Rbar-chi-A}
\end{equation}
which holds for any compact, boundaryless $\Sigma$.
For $\chi(\Sigma)\neq 0$, this relation can be inverted to yield
\begin{equation}
  A \ = \  \frac{4\pi\,\chi(\Sigma)}{\tilde R_{\Sigma}}\, ,
  \label{eq:A-chi-Rbar}
\end{equation}
so that, at fixed topology, specifying the area $A$ is algebraically equivalent to specifying the area-averaged intrinsic curvature $\tilde R_{\Sigma}$, and vice versa. We will exploit this one-to-one correspondence (for $\chi\neq 0$) to define a reference area by fixing a \emph{calibration} value of $\tilde R_{\Sigma}$, without assuming that ${}^{(2)}\!R$ is constant on the physical cross section.

For spherical sections one has $\chi=2$ and therefore $\tilde R_{\Sigma}>0$. For compact hyperbolic sections of genus $g\ge 2$, one has $\chi=2-2g<0$ and hence $\tilde R_{\Sigma}<0$, since $\tilde R_{\Sigma}=4\pi\chi(\Sigma)/A$.
In both cases, the ratio $\chi/\tilde R_{\Sigma}$ is positive, ensuring that equation~\eqref{eq:A-chi-Rbar} yields a positive area; constant-curvature metrics provide convenient canonical realizations of the normalization. The degenerate case is $\chi=0$ (torus), for which Gauss--Bonnet gives $\int_\Sigma \sqrt{\gamma}\,{}^{(2)}\!R\,d^2x=0$ and therefore $\tilde R_{\Sigma}=0$ independently of the area. Consequently, the Gauss--Bonnet theorem cannot fix a finite reference area at any non-zero calibration scale $|\tilde R_*|$. Including a toroidal sector would require an additional intrinsic datum beyond the Gauss--Bonnet average curvature, and is therefore not part of the minimal TCP calibration used below.

The TCP enforces the `no external scales' requirement by specifying a reference \emph{calibration datum} for the magnitude of the area-averaged intrinsic curvature, $|\tilde R_*|$. 
Through Gauss--Bonnet, this datum fixes a corresponding reference area in the cross-topology setting (and hence the area scale $A_*$ at which the entropy slope is evaluated).
This calibration datum should not be confused with $\tilde R_{\Sigma}$ of any particular physical horizon cross section; rather, it is a scheme choice that sets the normalization point for $s_0$.

In what follows, we keep $|\tilde R_*|$ fixed within each comparison (unless stated otherwise).
The absolute values in equation~\eqref{eq:A0-chi-TCP} below make the positivity of the reference area manifest for both spherical ($\chi>0$) and hyperbolic ($\chi<0$) topologies.
Fixing the calibration datum $|\tilde R_*|$ and using Gauss--Bonnet, one obtains for any compact representative with $\chi\neq 0$ a uniquely determined reference area,
\begin{equation}
  A_0(\chi)
  \ \equiv \  \frac{4\pi\,|\chi|}{|\tilde R_*|}\, ,
  \label{eq:A0-chi-TCP}
\end{equation}
which sets the evaluation area for the entropy slope in the cross-topology comparison.
Equivalently, $A_0(\chi)$ can be characterized as the unique area for which the \emph{area-averaged} intrinsic curvature of a compact representative of topology $\chi$ equals the chosen calibration scale $|\tilde R_*|$ (cf.\ equation~\eqref{eq:A-chi-Rbar}).
Constant-curvature metrics provide a convenient canonical realization of this normalization, since in that case ${}^{(2)}\!R$ is uniform and coincides with the average; however, the TCP does not assume that the physical horizon cross section satisfies ${}^{(2)}\!R = \text{const}$.
The TCP then provides a concrete calibration prescription for the reference area used to evaluate the entropy slope. In the cross-topology setting at fixed $|\tilde R_*|$, we adopt
\begin{equation}
  A_* \ \equiv \ A_0(\chi)\, ,
\end{equation}
so that different topologies at the same calibration datum are assigned different coarse-graining areas, and hence --- via the entropy slope --- different effective couplings whenever $\delta\neq 1$. One could more generally allow $A_* = C\,A_0(\chi)$ with a dimensionless constant $C$, but this would reintroduce a scheme dependence in the normalization of $G_{\rm eff}$; the minimal calibration sets $C=1$.

Evaluating the entropy slope~\eqref{eq:slope-delta} at $A_0(\chi)$ and inserting into~\eqref{eq:Geff-Astar} gives the topology-dependent effective coupling
\begin{equation}
  s_0(\chi)
  \ = \ \eta\ \! \delta\ \! (4G)^{-\delta}\,A_0(\chi)^{\delta-1}\, ,
  \qquad
  G_{\rm eff}(\chi)
  \ = \ \frac{1}{4s_0(\chi)}
  \ = \ \frac{G}{\eta\,\delta}\left(\frac{4G}{A_0(\chi)}\right)^{\delta-1}\, .
\end{equation}
Equation~\eqref{eq:A0-chi-TCP} can then be rearranged to yield
\begin{equation}
  G_{\rm eff}(\chi)
  \ = \ \frac{G}{\eta\,\delta}\left(\frac{G\,|\tilde R_*|}{\pi\,|\chi|}\right)^{\delta-1},
  \label{eq:Geff-chi}
\end{equation}
making explicit that, at fixed $|\tilde R_*|$,
\begin{equation}
  G_{\rm eff}(\chi)\ \propto\ |\chi|^{\,1-\delta}\, .
\end{equation}
Importantly, once the reference slope $s_0$ (equivalently, $A_*$) is fixed, the local Jacobson derivation is unchanged: on each patch the field equations remain
\begin{eqnarray}
G_{ab} \ = \ 8\pi G_{\rm eff}(\chi)\,T_{ab} \ + \ \Lambda g_{ab}\, ,
\end{eqnarray}
with $G_{\rm eff}$ determined by the entropy slope evaluated at the calibrated reference area. The TCP therefore introduces no additional local dynamics; it is a normalization prescription that fixes the evaluation scale entering $s_0$ (and hence $G_{\rm eff}$) in terms of intrinsic geometric data.
Although our explicit formulas have focused on the phenomenologically simple power-law branch~\eqref{eq:entropy-law-local}, the TCP logic relies only on the existence of a well-defined entropy slope $s_0 = \partial S/\partial A$ evaluated at the calibrated area. For a more general functional $S(A,\chi)$, potentially with explicit topological dependence, the slope $s_0$ computed at $A_* = A_0(\chi)$ will, in general, depend on both $A_*$ and $\chi$, resulting in a topology-dependent $G_{\rm eff}$ of the same structural form as in equation~\eqref{eq:Geff-chi}. In this sense, the TCP imposes a self-consistency condition on how generalized horizon entropies behave under admissible cross-topology comparisons. A universal area-type law cannot be chosen arbitrarily if the corresponding thermodynamic coupling is required to remain stable across compact sectors at fixed intrinsic curvature scale; any explicit topological dependence, if introduced, must be compatible with the Gauss--Bonnet calibration.

\subsection{Bounds from topology and scale}
\label{subsec:TCP-bounds}

Having established equation~\eqref{eq:Geff-chi}, one can distinguish two logically independent uses, which we will now discuss separately.

\subsubsection{{Fixed intrinsic scale, varying topology}}

Suppose that, for a given $|\tilde R_*|$, one compares horizon cross--sections with different Euler characteristic $\chi_1,\chi_2$. With universal $(\eta,\delta)$ the TCP predicts
\begin{equation}
  \frac{G_{\rm eff}(\chi_1)}{G_{\rm eff}(\chi_2)}
  \ = \  \left(\frac{|\chi_2|}{|\chi_1|}\right)^{\delta-1}\, .
  \label{eq:Geff-chi-ratio}
\end{equation}
This expresses a purely topological modulation of the locally inferred coupling at fixed intrinsic curvature scale, with $A_* = A_0(\chi)$ at fixed $|\tilde R_*|$.

As a concrete illustration, consider four--dimensional AdS black holes with horizons of constant curvature $k=\pm 1$
\begin{equation}
  ds^2 \ = \  -f(r)\,dt^2 \ + \  \frac{dr^2}{f(r)} \ + \  r^2\,d\Sigma_k^2\, ,
  \qquad
  f(r) \ = \  k \ - \  \frac{2GM}{r} \ + \  \frac{r^2}{L^2}\, ,
\end{equation}
where $d\Sigma_k^2$ is the metric on a compact two--manifold $\Sigma_k$ with Euler characteristic $\chi(\Sigma_k)$. Typical examples are $\chi=2$ for the sphere ($k=+1$) and $\chi=2-2g<0$ for higher--genus hyperbolic horizons ($k=-1$), obtained as compact quotients of constant--curvature spaces. At the horizon $r = r_h$, defined as the largest positive root of $f(r)=0$, the intrinsic scalar curvature of $\Sigma_k$ is
\begin{equation}
  {}^{(2)}R \ = \ \frac{2k}{r_h^2}\, ,
\end{equation}
so that constant--curvature spherical and hyperbolic horizons with the same radius $r_h$ share the same intrinsic scale $|\tilde R_{\Sigma}|=|{}^{(2)}R|=2/r_h^2$ (for constant--curvature horizons). 
Evaluating the TCP at a calibration scale chosen to match this common value, $|\tilde R_*|=2/r_h^2$, Gauss--Bonnet then implies
\begin{equation}
  A_0(\chi) \ = \  \frac{4\pi\,|\chi|}{|\tilde R_*|}
            \ = \  2\pi\,|\chi|\,r_h^2\, ,
\end{equation}
and the ratio \eqref{eq:Geff-chi-ratio} specializes to
\begin{equation}
  \frac{G_{\rm eff}(\chi_{\rm hyp})}{G_{\rm eff}(\chi_{\rm sph})}
  \ = \  \left(\frac{|\chi_{\rm sph}|}{|\chi_{\rm hyp}|}\right)^{\delta-1}
  \ =  \ \left(\frac{2}{2g-2}\right)^{\delta-1}\, ,
\end{equation}
for a hyperbolic horizon of genus $g\ge2$ compared to a spherical one.

If independent information (for instance, from bulk dynamics or holographic data) indicates that the effective coupling should not differ by more than a fractional tolerance $\Delta\in[0,1)$ between two topologies at the same intrinsic scale, one can impose
\begin{equation}
  (1+\Delta)^{-1}\ \le\ 
  \frac{G_{\rm eff}(\chi_1)}{G_{\rm eff}(\chi_2)}
  \ \le\ 1+\Delta\, .
\end{equation}
Using~\eqref{eq:Geff-chi-ratio}, and the monotonicity of the logarithm, this is equivalent to the logarithmic bound
\begin{equation}
  |1-\delta|\ \le\ 
  \frac{\ln(1 \ + \ \Delta)}
       {\big|\ln(|\chi_1|/|\chi_2|)\big|}\, .
  \label{eq:delta-bound-TCP}
\end{equation}
In~\eqref{eq:delta-bound-TCP} we have kept $|1-\delta|$ so that the bound is valid for both branches $\delta<1$ and $\delta>1$ and for either ordering of $|\chi_1|$ and $|\chi_2|$. For the weakly non--extensive branch $\delta\le 1$ relevant here one simply has $|1-\delta|=1-\delta$, so the TCP constraint~\eqref{eq:delta-bound-TCP} reduces to the one-sided bound
\begin{equation}
  1 \ - \ \delta\ \le\ 
  \frac{\ln(1 \ + \ \Delta)}
       {\big|\ln(|\chi_1|/|\chi_2|)\big|}\, .
\end{equation}

This shows that, in the sub-area branch $\delta\leq1$ emphasized here, any stability requirement imposed on sectors with distinct, non-zero values of $|\chi|$ at fixed intrinsic curvature scale forces $\delta$ to approach the Bekenstein--Hawking value from below. More generally, the symmetric bound~\eqref{eq:delta-bound-TCP} constrains both branches by forcing $|1-\delta|$ to be small whenever such sectors are admitted by the underlying theory or boundary conditions and can be consistently compared.

The bound~\eqref{eq:delta-bound-TCP} applies only to sectors with non-zero Euler characteristic and with different values of $|\chi|$. As discussed in section~\ref{subsec:TCP-minimal}, the toroidal case ($\chi=0$) is degenerate for the present Gauss--Bonnet calibration: it does not yield a finite reference area at fixed non-zero $|\tilde R_*|$, leaving the logarithmic ratio undefined. Such a sector would require an additional intrinsic datum and therefore lies outside the minimal no-external-scale TCP comparison.

For illustration, consider a comparison between a spherical section ($|\chi_{\rm sph}|=2$) and a genus-$3$ hyperbolic compact quotient ($|\chi_{\rm hyp}|=4$). This provides the simplest hyperbolic example that leads to a non-trivial modification relative to the spherical case. Since the minimal TCP bound depends on $|\chi|$, a genus-$2$ quotient has $|\chi|=2$ and therefore gives a unit ratio when compared with the spherical case, yielding no logarithmic constraint.
Requiring 
$\big|G_{\rm eff}(\chi_{\rm hyp})/G_{\rm eff}(\chi_{\rm sph})-1\big|\lesssim 10\%$
(i.e., $\Delta=0.1$) then yields
\begin{eqnarray}
  1 \ - \ \delta
  \ \lesssim\ 
  \frac{\ln(1.1)}{\ln 2}
  \ \approx\ 0.14\, ,
\end{eqnarray}
so $\delta\gtrsim 0.86$ in the branch $\delta\le1$ considered here. A still modest genus-$5$ example, for which $|\chi_{\rm hyp}|=8$ and therefore the logarithmic factor is $\ln(|\chi_{\rm hyp}|/|\chi_{\rm sph}|)=\ln(8/2)=\ln4$, gives
\begin{eqnarray}
  1-\delta
  \ \lesssim\
  \frac{\ln(1.1)}{\ln 4}
  \ \approx \ 7\times 10^{-2}\, ,
\end{eqnarray}
that is, $\delta\gtrsim 0.93$ for the same $10\%$ tolerance.

More generally, for a spherical section compared with a genus-$g$ hyperbolic compact quotient, equation~\eqref{eq:delta-bound-TCP} gives
\begin{equation}
  |1-\delta|
  \ \le\
  \frac{\ln(1+\Delta)}{\ln(g-1)}\, ,
  \qquad (g\ge3)\, .
\end{equation}
Thus, the topological strengthening of the bound is logarithmic and controlled by the admissible range of genera, rather than by any single preferred numerical choice. These estimates should therefore be understood as scaling estimates within theories or boundary conditions that admit such compact sectors, not as astrophysical probes for ordinary four-dimensional asymptotically flat general relativity.

It is nevertheless important to keep the status of this `purely topological' lever in proper perspective. In ordinary four-dimensional asymptotically flat general relativity, stationary black-hole horizons are topologically restricted by the standard topology and rigidity theorems, while all operationally accessible cosmological horizons appear to possess spherical topology. Thus, in the directly observed Universe, effectively only the spherical case $\chi=2$ is tested. The topological part of the TCP should therefore be read primarily as a consistency constraint on any non-spherical sectors admitted by the chosen theory or boundary conditions and compared at fixed intrinsic curvature scale, rather than as a direct observational constraint.
It quantifies how strongly non-extensive exponents would be restricted \emph{if} admissible families of horizons with sufficiently separated non-zero Euler characteristics were physically realized and if their effective couplings could be compared at fixed intrinsic curvature scale. Within such scenarios, equation~\eqref{eq:delta-bound-TCP} provides a clean bound: once $\Delta$ and the available range of $|\chi|$ are specified, sizable deviations from the Bekenstein--Hawking value become incompatible with an emergent-gravity interpretation based on horizon thermodynamics.

In holographic contexts, the TCP acquires a particularly sharp interpretation. Within AdS/CFT, the bulk Newton constant governing horizon entropy simultaneously determines the central charge (or, equivalently, the stress--tensor two-point coefficient) of the dual CFT. At fixed AdS radius, any topology dependence of $G_{\rm eff}(\chi)$ at the horizon would therefore induce a corresponding variation of the CFT central charge across admissible topological sectors. The absence of such a variation in a given holographic construction can thus be rephrased, in our framework, as a nontrivial constraint on $(\eta,\delta)$: consistency between the bulk TCP and boundary universality of the central charge strongly constrains $\delta$ toward $1$, unless the assumption that a single non-extensive exponent governs all admitted horizon topologies breaks down.

\subsubsection{Fixed topology, running intrinsic scale} 

As a complementary application of the TCP, we now consider a fixed topology (e.g., $\chi=\chi_0$) and interpret the dependence of $G_{\rm eff}(A_*)$ on the reference area as a controlled scale dependence of the coarse-grained coupling induced by variations of the horizon size within the same topological class. From equation~\eqref{eq:Geff-Astar}, and assuming universal $(\eta,\delta)$, one finds
\begin{equation}
  \frac{G_{\rm eff}(A_{*2})}{G_{\rm eff}(A_{*1})}
  \ = \  \left(\frac{A_{*1}}{A_{*2}}\right)^{\delta-1}\, .
  \label{eq:Geff-scale-ratio}
\end{equation}
In this application, one no longer compares different topologies: the TCP is implemented within a single topological class (fixed $\chi$), and the relevant reference area is identified with the \emph{physical} horizon area of the screen under consideration, $A_*\equiv A$ (e.g., $A\sim r_h^2$ for AdS black holes), rather than with the Gauss--Bonnet reference area $A_0(\chi)$ introduced for cross-topology comparisons at fixed intrinsic scale. The underlying mechanism remains unchanged --- the local coupling is still determined by the entropy slope as $G_{\rm eff}(A_*)=1/[4\,s_\delta(A_*)]$ --- but the identification of $A_*$ now tracks variations in the horizon size within the same topological sector. Any deviation from $\delta=1$ therefore manifests as a controlled scale dependence of the effective coupling between, for example, small and large AdS black holes of the same topology.

In the minimal sector analyzed below, the exponent $\delta$ is held fixed. This means that the entropy law is the same throughout the comparison, while the evaluation area $A_*$ changes with the physical horizon scale. A more general scale-dependent entropy response can be parametrized without committing to a specific functional form $S(A)$ by defining a local logarithmic running index directly through the slope-induced coupling,
\begin{equation}
  \frac{d\ln G_{\rm eff}}{d\ln A_*}
  \ \equiv\ 1-\delta_{\rm eff}(A_*)\, .
  \label{eq:delta-eff-def}
\end{equation}
This definition is deliberately made at the level of $G_{\rm eff}$, or equivalently of the entropy slope, because a direct replacement $\delta\to\delta(A_*)$ inside the power law $S(A)=\eta(A/4G)^\delta$ would also generate derivative terms from $d\delta/dA_*$. The constant-$\delta$ model is recovered when $\delta_{\rm eff}(A_*)=\delta$. Integrating equation~\eqref{eq:delta-eff-def} gives
\begin{equation}
  \ln\frac{G_{\rm eff}(A_{*2})}{G_{\rm eff}(A_{*1})}
  \ =\
  \int_{\ln A_{*1}}^{\ln A_{*2}}
  \big[1-\delta_{\rm eff}(A)\big]\,d\ln A\, .
  \label{eq:running-delta-integral}
\end{equation}
Thus, in a running-index extension, stability bounds constrain the logarithmic average of $1-\delta_{\rm eff}$ over the interval of areas being compared. In the following numerical estimates we restrict to the constant-index specialization, which is the minimal one-parameter sector studied in this work.

In practice, the observed near constancy of Newton's constant over a wide range of curvature scales can be implemented by requiring that the effective coupling $G_{\rm eff}$ inferred from horizon thermodynamics vary by at most a fractional amount $\Delta_{\rm run}$ between two physically relevant scales. Imposing this condition on equation~\eqref{eq:Geff-scale-ratio} then yields
\begin{equation}
  (1+\Delta_{\rm run})^{-1}
  \ \le\ 
  \frac{G_{\rm eff}(A_{*2})}{G_{\rm eff}(A_{*1})}
  \ \le\ 
  1+\Delta_{\rm run}\, ,
\end{equation}
and hence
\begin{equation}
  |1-\delta|
  \;\le\;
  \frac{\ln(1+\Delta_{\rm run})}{\big|\ln(A_{*2}/A_{*1})\big|}\, .
  \label{eq:delta-bound-scale}
\end{equation}

For a running entropy index, the same stability requirement constrains the logarithmic average of $1-\delta_{\rm eff}(A)$ over the interval $[A_{*1},A_{*2}]$, as in equation~\eqref{eq:running-delta-integral}. The explicit bound~\eqref{eq:delta-bound-scale} is therefore the constant-index specialization of this more general integral constraint.

It is useful to make explicit how the constraint depends on the available dynamical range in the horizon area. Assuming $A_{*2}>A_{*1}$, equation~\eqref{eq:delta-bound-scale} can be rewritten as
\begin{equation}
  |1-\delta|
  \ \le\ 
  \frac{\ln(1+\Delta_{\rm run})}{\ln\!\left(A_{*2}/A_{*1}\right)}\,,
\end{equation}
which shows that, for a fixed tolerance $\Delta_{\rm run}$, the bound tightens only logarithmically as the ratio $A_{*2}/A_{*1}$ increases. For a representative choice $\Delta_{\rm run}=0.1$ and a moderate dynamical range $A_{*2}/A_{*1}=100$ (e.g., two spherical AdS horizons with $r_{h,2}/r_{h,1}=10$), one finds
\begin{eqnarray}
  |1-\delta|
  \ \lesssim \ 
  \frac{\ln(1.1)}{\ln(100)}
  \;\approx\;2\times 10^{-2}\, ,
\end{eqnarray}
so that $\delta \gtrsim 0.98$ within our working non-extensive window. Larger area ratios (e.g., connecting stellar-mass and supermassive black-hole horizons within the same entropy law) tighten the constraint according to the same logarithmic scaling.

A more extreme --- and cosmologically well-motivated --- dynamical range within the same topological class (spherical sections, $\chi=2$) arises when cosmological horizons are incorporated into the fixed-topology scale-dependence analysis. In this extension, the TCP mechanism remains unchanged: $G_{\rm eff}$ is determined by the entropy slope evaluated at a resolution scale $A_*$, which is now allowed to vary over many orders of magnitude. The only additional physical input is the identification of the appropriate thermodynamic screen in a dynamical spacetime.

For the FRW universe, the standard local causal screen used in Clausius-type derivations is the \emph{apparent horizon}, defined as the marginally trapped surface on which the expansion of one of the radial null congruences vanishes, thereby separating trapped and untrapped regions. It admits a geometric first-law structure closely analogous to that of black holes~\cite{Hayward1998}. 
Its radius is
\begin{equation}
  R_A(a) \ = \ \frac{1}{\sqrt{H(a)^2+k/a^2}} \ \simeq \ H(a)^{-1}\, ,
\end{equation}
where the final approximation assumes $k\simeq 0$. The corresponding area is therefore
\begin{equation}
  A_H(a) \ = \ 4\pi R_A^2 \ \simeq \  \frac{4\pi}{H(a)^2}\, ,
  \label{eq:AH-H}
\end{equation}
as in the standard apparent-horizon thermodynamic derivations of Friedmann dynamics \cite{Hayward1998,CaiKim2005}.
Identifying the reference area with the apparent-horizon area, $A_*\equiv A_H$, equation~\eqref{eq:Geff-Astar} gives
\begin{equation}
  G_{\rm eff}(a)
  \ = \  \frac{G}{\eta \delta}\left(\frac{4G}{A_H(a)}\right)^{\delta-1}
 \  \propto  \ A_H(a)^{\,1-\delta}
  \ \propto \ H(a)^{2(\delta-1)}\,  ,
  \label{eq:Geff-H}
\end{equation}
where the omitted prefactor is $(G/(\eta\delta))\,(G/\pi)^{\delta-1}$ in the $k\simeq 0$ approximation, using $A_H(a)\simeq 4\pi/H(a)^2$. 
Since our focus is on the scale dependence of $G_{\rm eff}$, it is convenient to remove the scheme-dependent prefactor by working with the dimensionless modulation normalized to unity today
\begin{equation}
  \mu(a)\  \equiv \ \frac{G_{\rm eff}(a)}{G_{\rm eff}(1)}
  \ = \ \left[\frac{A_H(a)}{A_H(1)}\right]^{1-\delta}
  \ = \ \left[\frac{H(a)}{H_0}\right]^{2(\delta-1)}\, .
  \label{eq:mu-a}
\end{equation}
Under the universality assumption for $(\eta,\delta)$ and the identification $A_*=A_H$, the TCP therefore fixes a definite redshift dependence of the effective gravitational coupling for a given background expansion history $H(a)$. In this minimal framework, the time dependence of $\mu(a)$ is fully determined by $H(a)$, while the exponent is set by the constant non-extensivity index $\delta$ of the entropy law.

The running-index generalization introduced in equations~\eqref{eq:delta-eff-def}--\eqref{eq:running-delta-integral} can be specialized to the cosmological setting by identifying $A_*=A_H(a)$, in which case the effective index may be viewed equivalently as $\delta_{\rm eff}(A_H(a))$ or $\delta_{\rm eff}(z)$. We do not analyze this enlarged functional freedom here. The cosmological estimates and growth curves below refer to the minimal constant-$\delta$ model, for which equation~\eqref{eq:mu-a} gives the one-parameter prediction.

A first estimate follows by applying the scale-running bound~\eqref{eq:delta-bound-scale} to a deliberately conservative comparison between the present cosmological apparent-horizon area and a supermassive black-hole horizon area.
Taking $R_{A0}\equiv R_A(a{=}1)\sim H_0^{-1}$ and $R_s\sim 2GM$ (up to $\mathcal{O}(1)$ spin-dependent corrections, which are irrelevant at this level of accuracy), one finds $A_H/A_{\rm BH}\sim (R_{A0}/R_s)^2\sim 10^{27\text{--}28}$ at order of magnitude, corresponding to $\log_{10}(A_H/A_{\rm BH})\sim 27$--$28$.
Imposing a fractional stability $|\Delta G/G|\lesssim \Delta_{\rm run}$ across this range yields
\begin{equation}
  |1-\delta|
  \ \lesssim\
  \frac{\ln(1+\Delta_{\rm run})}{\big|\ln(A_H/A_{\rm BH})\big|}
  \ \sim\ 10^{-3}\, ,\qquad \text{for}\ \ \Delta_{\rm run}\sim 10^{-1}\, .
  \label{eq:delta-bound-cosmobH}
\end{equation}
This estimate should be interpreted as a \emph{consistency filter} under the working identification of the thermodynamic coupling with the effective gravitational strength constrained by astrophysical and cosmological observations.  For percent-level tolerances, $\Delta_{\rm run}\sim 10^{-2}$ --- as suggested by existing bounds on $|\Delta G/G|$ from early- and late-universe consistency tests --- the constraint tightens to $|1-\delta|\lesssim 2\times 10^{-4}$, see, e.g.,~\cite{CosmoVaryingG1}. In particular, benchmark values such as the microcanonical exponent $\delta=3/2$ would imply $G_{\rm eff}(A_H)/G_{\rm eff}(A_{\rm BH})\sim (A_H/A_{\rm BH})^{1/2}\sim 10^{13\text{--}14}$, incompatible by many orders of magnitude unless the universality assumption on $\delta$ is abandoned.

More specifically, the constant-index scale dependence encoded in \eqref{eq:mu-a} becomes observationally testable once propagated into the linear growth of structure. 
The resulting imprint on $f\sigma_8(z)$ exhibits a characteristic redshift-dependent suppression or enhancement (for $\delta<1$ or $\delta>1$, respectively) that is rigidly tied to the background expansion through~\eqref{eq:mu-a}, rather than introduced as an arbitrary function. In the quasi-static approximation for linear sub-horizon modes ($k\gg aH$), and neglecting scale-dependent stresses beyond this effective coupling description, a broad class of scale-independent modified-gravity parameterizations enters the growth equation through $\mu(a)$ \cite{Linder2005,AmendolaTsujikawa2010}. Denoting by $D(a)$ the linear growth factor, $f(a)=d\ln D/d\ln a$ the growth rate, and $\sigma_8(a)\propto D(a)$ the fluctuation amplitude on $8\,h^{-1}{\rm Mpc}$, one considers the observable combination $f\sigma_8(z)$. To isolate this constant-index TCP modification of the effective coupling, we keep the background expansion fixed to flat $\Lambda$CDM and propagate departures from GR only through $\mu(a)$.
We set the gravitational slip to zero and encode departures from GR solely through the effective Poisson strength $\mu(a)$. 
In the quasi-static, sub-horizon regime for pressureless matter, the resulting growth equation reads
\begin{equation}
  D''(a)
  \ + \ \left[\frac{3}{a} \ + \ \frac{H'(a)}{H(a)}\right]D'(a)
  \ - \ \frac{3}{2}\frac{\Omega_m(a)}{a^2}\,\mu(a)\,D(a)\ = \ 0\, ,
  \label{eq:growth-mu}
\end{equation}
where the prime denotes $d/da$, and $\Omega_m(a)=\Omega_{m0}H_0^2/(a^3H(a)^2)$ is the standard matter fraction of the assumed $\Lambda$CDM background.
\begin{figure}[tbp]
  \centering
  \includegraphics[width=0.85\textwidth]{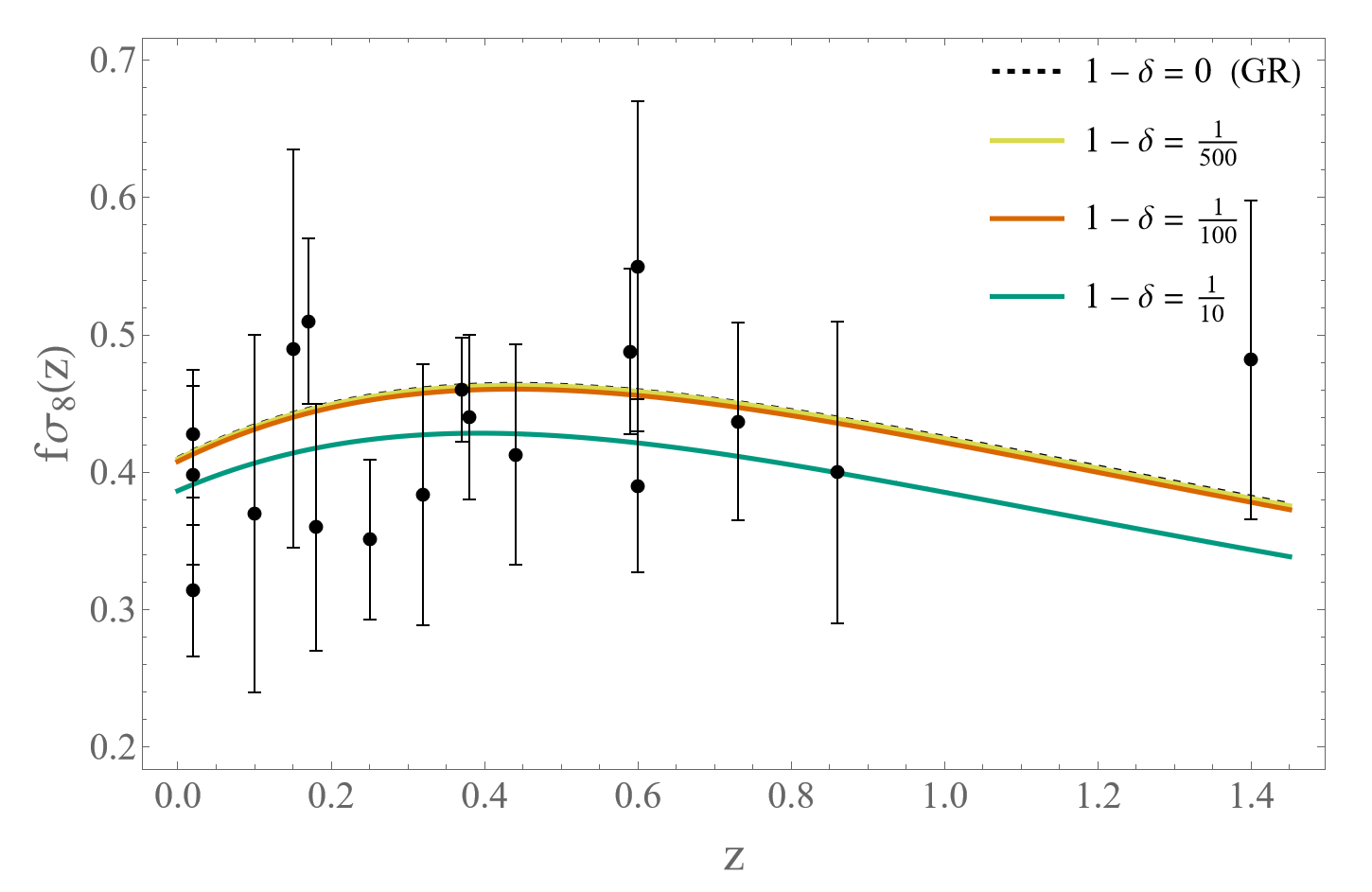}
 \caption{
    TCP-induced scale dependence of the effective coupling (at fixed topology) and its impact on the growth observable $f\sigma_8(z)$.
    The reference area is identified with the cosmological apparent-horizon area, $A_*\equiv A_H=4\pi R_A^2$, where $R_A=(H^2+k/a^2)^{-1/2}$; hence $A_H=4\pi/H^2$ for $k=0$, see, e.g.,~\cite{Hayward1998,CaiKim2005}. Using the entropy--slope relation~\eqref{eq:Geff-Astar}, one obtains the dimensionless modulation $\mu(a)\equiv G_{\rm eff}(a)/G_{\rm eff}(1)=[H(a)/H_0]^{2(\delta-1)}$, cf.~\eqref{eq:mu-a}. The curves show $f\sigma_8(z)$ computed by integrating~\eqref{eq:growth-mu} on a fixed flat $\Lambda$CDM background with $(\Omega_{m0},\Omega_{\Lambda 0})=(0.3,0.7)$ (neglecting radiation at the redshifts of interest), growth normalization $D(1)=1$, and $\sigma_8=0.8$, for $\delta=1$ (GR) and for $|1-\delta|\in\{1/500,\;1/100,\;1/10\}$. Points with $1\sigma$ uncertainties are taken from the compilation of $f\sigma_8$ measurements in~\cite{NesserisPantazisPerivolaropoulos2017} and are shown for comparison only (no fit is performed). The figure refers to the constant-$\delta$ minimal model; a running entropy index would correspond to the generalized logarithmic prescription discussed around equation~\eqref{eq:running-delta-integral}.}
  \label{fig:tcp-fsigma8}
\end{figure}
In the constant-$\delta$ minimal model, figure~\ref{fig:tcp-fsigma8} illustrates this behavior for representative values of $|1-\delta|$ and overlays a standard compilation of current measurements for \cite{NesserisPantazisPerivolaropoulos2017}.
This provides a direct connection between the TCP framework and dark-sector phenomenology. In particular, the TCP does not introduce any additional dark-matter component; instead, it modifies the gravitational response through the entropy-slope relation, leading to a redshift-dependent effective coupling $G_{\rm eff}(a)$. In this sense, the resulting dynamics can be interpreted as an effective modification of the dark sector, whose imprint appears in the same observables used to probe dark matter and modified gravity, including structure growth, weak lensing, and large-scale clustering.

A comparison of the specific redshift scaling~\eqref{eq:mu-a} with survey likelihoods would require incorporating the TCP prediction into the growth-sector parameterizations used in full-shape and weak-lensing analyses, including a consistent treatment of bias and projection effects. Forecasts for upcoming surveys indicate percent-level sensitivity to scale-independent modifications of the growth rate over $0\lesssim z\lesssim 2$, making \eqref{eq:mu-a} a concrete target for next-generation large-scale-structure data~\cite{EuclidPrepVII2020,EuclidPrepXV2022}. In summary, applying the TCP scale-dependence framework to cosmology yields two main results: a stringent consistency bound~\eqref{eq:delta-bound-cosmobH}, which requires any \emph{universal} non-extensive exponent to lie very close to the Bekenstein--Hawking value $\delta=1$, and a falsifiable growth signature~\eqref{eq:mu-a}--\eqref{eq:growth-mu}, whose characteristic redshift dependence can be tested by current and upcoming surveys.

\subsection{Status and variants of the TCP}
\label{subsec:TCP-status}

A different logical requirement (still at fixed $|\tilde R_*|$) is to demand that the local coupling be independent of topology, i.e., 
\begin{equation}
  G_{\rm eff}(\chi) \ \equiv \  G_{\rm phys}\, ,
\end{equation}
for all compact sectors admitted by the chosen theory or boundary conditions, for instance spherical sections and higher-genus hyperbolic quotients in settings where both types of sectors are available.
With universal $(\eta,\delta)$ this immediately forces $\delta=1$, so that the non--extensive branch collapses back to the Bekenstein--Hawking law. If one wishes to retain $\delta \neq 1$ while enforcing $G_{\rm eff} = G_{\rm phys}$, the universality of $\eta$ must be relaxed and allowed to depend on topology.  Solving~\eqref{eq:Geff-chi} for $\eta$ at fixed $\delta$ and $G_{\rm phys}$ yields
\begin{equation}
  \eta(\chi)
  \ = \ \frac{G}{\delta\,G_{\rm phys}}
   \left(\frac{G\,|\tilde R_*|}{\pi\,|\chi|}\right)^{\delta-1}.
  \label{eq:eta-chi-TCP}
\end{equation}
In this variant --- non-extensive at the level of the entropy functional but with a universal effective coupling --- the Jacobson construction on each local horizon patch still yields Einstein equations with coupling $G_{\rm phys}$. However, the global entropy prefactor is no longer universal and instead acquires explicit dependence on topology.
A simple illustration of this logic is provided by Einstein gravity supplemented by a four-dimensional Gauss--Bonnet term. In this case, the Gauss--Bonnet density is topological and does not affect the bulk equations of motion, while the Wald entropy acquires an additive contribution of the form $S_{\rm GB} \propto \alpha\,\chi(\Sigma)$. This term is independent of the horizon area, with $dS_{\rm GB}/dA = 0$, and therefore acts as a purely topological contribution that can be used to calibrate the overall normalization of $S(A)$ without modifying the local field equations. The corresponding algebraic relations between $(\eta,\delta)$ and the coupling $\alpha$ are collected in appendix.

From a conceptual perspective, the TCP plays a dual role. First, it implements a concrete `no new scale' principle: once an intrinsic curvature scale $|\tilde R_*|$ is fixed, the calibration scale is determined entirely by intrinsic horizon data through the Gauss--Bonnet theorem, without reference to external length scales or arbitrary renormalization areas. Second, it promotes horizon scale and, where distinct compact sectors are admissible, topology to quantitative consistency tests for non-extensive, area-type entropies: different choices of $(\eta,\delta)$ lead to definite predictions for the scale dependence of $G_{\rm eff}$ and, when such sectors are admitted by the chosen theory or boundary conditions, for its topology dependence. These predictions can then be confronted with theoretical consistency requirements or observational constraints.

It is therefore useful to summarize explicitly the hypothesis space within which our quantitative bounds should be interpreted. 
\begin{enumerate}
  \item  We assume that horizon entropies are of area type, $S(A)=\eta(A/4G)^{\delta}$, with a well-defined slope $s_0=\partial S/\partial A$ that determines a thermodynamic coupling $G_{\rm eff}=1/(4s_0)$ as in equation~\eqref{eq:Geff-Astar}. 
  \item When comparing different topologies at fixed intrinsic curvature scale, we impose the TCP calibration $A_*=A_0(\chi)$, thereby implementing the `no external scales' requirement discussed in section~\ref{subsec:TCP-minimal}; for scale-running arguments at fixed topology, we instead identify $A_*$ with the physical horizon area. 
  \item  Whenever $G_{\rm eff}$ is confronted with data, we adopt the working hypothesis  that the thermodynamic coupling inferred from local horizon patches can be identified, up to order-unity scheme ambiguities, with the effective Newton constant $G_{\rm phys}$ constrained by laboratory, astrophysical, and cosmological measurements.
\end{enumerate}
Within this framework, the TCP introduces no additional dynamics beyond the local Jacobson construction; rather, it acts as a calibration principle that renders non-extensive area-type proposals quantitatively testable. Once $(\eta,\delta)$ are specified, the horizon scale and, where admissible, the horizon topology feed into $G_{\rm eff}$ through the Gauss--Bonnet calibration and the Clausius relation. The resulting variation of the effective coupling across horizon scales and, where physically meaningful, across topological sectors then becomes a sharp target for both microscopic models and phenomenological bounds.

\section{Conclusion and outlook}
\label{sec:conclusion}

In this paper, we combined non-extensive and, more generally, non-additive horizon thermodynamics with Jacobson's emergent-gravity paradigm to demonstrate how generalized area-law entropies can be consistently incorporated while preserving a fully thermodynamic origin of the gravitational field equations. We took as a starting point the view that the entropy functional $S(A)$ is not a microscopic entanglement measure, but rather an effective, coarse-grained constitutive relation for horizon thermodynamics, encoding long-range correlations and/or long-distance interactions that lie beyond the conventional Boltzmann--Gibbs paradigm. Within this framework, gravity emerged from thermodynamic consistency conditions imposed on local Rindler horizons. The analysis naturally separated into two complementary levels: a local level, where the field equations followed in canonical form from horizon thermodynamics on Rindler patches, and a calibration level, where intrinsic geometric data and, when distinct compact sectors are admitted by the chosen theory or boundary conditions, topological data fixed the reference scale entering the local construction and thereby determined the effective gravitational coupling.

At the local level, we recast Jacobson's horizon argument as the stationarity condition of a Massieu functional at fixed Unruh temperature. In the area-type sector --- where the horizon entropy density was taken to be uniform on the local patch and determined by the slope $s_0 = \left.dS/dA\right|_{A_*}$ of the phenomenological power law $S(A)=\eta(A/4G)^{\delta}$ --- the canonical balance reproduced Einstein's field equations with an effective Newton coupling $G_{\rm eff}=1/(4s_0)$.  

As a separate non-equilibrium scalar-response extension, we also showed that the same Massieu language accommodates a local, non-constant horizon entropy density. The familiar field equations of $f(R)$ gravity are recovered only after matching this macroscopic response to the Wald entropy density, $\Phi=f'(R)$, and supplementing the Clausius relation by internal entropy production. This scalar-response extension is logically separate from the area-type branch used in the TCP bounds. In the minimal sector relevant for the TCP, the local horizon balance probes only the entropy slope evaluated at the chosen reference area $A_*$, within a fixed thermodynamic scheme (fixed $\beta$, fixed $\rho$, and fixed evaluation scale during $\delta_g$).

The remaining question was how to fix the reference scale $A_*$ without introducing external macroscopic scales. This was addressed by the Topological Calibration Principle. For compact, boundaryless horizon cross-sections with $\chi\neq0$ admitted by the chosen theory or boundary conditions, the Gauss--Bonnet theorem relates the area, the area-averaged intrinsic curvature, and the Euler characteristic. Fixing a calibration value $|\tilde R_*|$ then selects, for each admissible topological class, a canonical reference area $A_0(\chi)$; constant-curvature metrics provide a convenient realization of this normalization. In cross-topology comparisons at fixed $|\tilde R_*|$, the TCP identifies $A_* \equiv A_0(\chi)$. 

Importantly, the TCP does not introduce non-local dynamics into the field equations; rather, it provides a principled normalization --- effectively a gauge fixing of the evaluation scale $A_*$ --- for the single macroscopic input probed locally, namely the entropy slope. For the non-extensive law $S(A)=\eta(A/4G)^\delta$, this implies a topology-dependent effective coupling $G_{\rm eff}(\chi)\propto|\chi|^{\,1-\delta}$ within an admissible family of compact sectors, while at fixed topology the same slope mechanism yields a controlled scale dependence $G_{\rm eff}(A_*)$.

These two applications yielded explicit logarithmic constraints on $|1-\delta|$. At fixed intrinsic curvature scale, comparing two admissible Euler characteristics led to the topological bound~\eqref{eq:delta-bound-TCP}; at fixed topology, comparing different horizon areas led to the scale-running bound~\eqref{eq:delta-bound-scale}. Under the assumptions of universality of $(\eta,\delta)$ and of an identification between the local thermodynamic coupling and the effective gravitational coupling constrained by data, these results act as a stringent selection rule: only entropy exponents that keep $G_{\rm eff}$ sufficiently stable across large area ranges and, where physically meaningful, across admissible topological sectors remain compatible with an emergent-gravity interpretation based on local horizon thermodynamics. In particular, benchmark values suggested by microcanonical group-entropy arguments, such as $\delta=3/2$, would generically imply order-of-magnitude variations of $G_{\rm eff}$ across horizon scales, and are therefore excluded within this universal-sector reading unless universality is relaxed or the mapping between $G_{\rm eff}$ and the Newtonian limit is modified.

A notable byproduct of the TCP was that it provided concrete, falsifiable signatures rather than merely consistency bounds. In particular, in the cosmological framework, when the apparent-horizon area was identified with the reference evaluation scale for the entropy slope, $A_* \equiv A_H \simeq 4\pi/H^2$ in the spatially flat case, the constant-index sector yielded a fixed redshift dependence for the normalized coupling modulation,
$\mu(a)=G_{\rm eff}(a)/G_{\rm eff}(1)=[H(a)/H_0]^{2(\delta-1)}$, cf.~equation~\eqref{eq:mu-a}.
Employing this prediction in the standard linear growth equation produced a characteristic imprint on $f\sigma_8(z)$, shown in figure~\ref{fig:tcp-fsigma8}. The essential point was structural: within the TCP framework, the growth-sector modification was not an arbitrary function but a minimal, one-parameter deformation whose redshift dependence was fully determined by the observed expansion history. This made the constant-index TCP prediction directly testable against current and forthcoming large-scale-structure constraints.
This observation also places the present framework in direct contact with dark-sector phenomenology. The TCP should not be interpreted as a model of particle dark matter; rather, it provides a thermodynamic mechanism that modifies the effective gravitational coupling in a controlled way. Its connection with dark energy is more direct: within Jacobson-type constructions, the cosmological constant arises as an integration constant, while the cosmological apparent horizon provides the natural thermodynamic screen. In this setting, non-extensive horizon entropy can induce a minimal and testable deformation of the effective dark-energy sector, without introducing additional (exotic) degrees of freedom at the level of the action.

From a microscopic perspective, the TCP can be interpreted as a macroscopic compatibility requirement. Any candidate quantum-gravity mechanism yielding a generalized horizon entropy should not only reproduce an admissible local slope $dS/dA$ consistent with Jacobson's balance relation, but also remain compatible with the near invariance of the corresponding thermodynamic coupling across the large range of horizon scales and, where such sectors are admitted, across topological sectors. When such sectors are admitted by the underlying theory and boundary conditions and compared at fixed intrinsic curvature scale, topology is not merely a descriptive label: through the Gauss--Bonnet relation, it becomes a quantitative probe of how generalized entropies can encode intrinsic geometric information without reintroducing external scales.

Several natural extensions suggest themselves. First, it would be worthwhile to incorporate the TCP prediction for $\mu(a)$ as a minimal, one-parameter extension of $\Lambda$CDM into full cosmological inference frameworks --- including weak-lensing and full-shape analyses --- and to confront it with current and forthcoming large-scale-structure datasets (including DESI- and Euclid-class measurements), thus quantifying the sensitivity to $|1-\delta|$ under realistic nuisance modeling. A natural generalization would be to relax the constant-index assumption and constrain a running logarithmic entropy index $\delta_{\rm eff}(A_*)$ or $\delta_{\rm eff}(z)$ directly from data.
Second, holographic frameworks provide a particularly intriguing theoretical arena: since the bulk Newton constant determines boundary central charges, any TCP-induced change of $G_{\rm eff}$ across admissible horizon topologies should leave a corresponding boundary imprint, thereby promoting topology to a precision consistency test in settings where distinct topological sectors are part of the theory.
Third, extending the calibration logic beyond $f(R)$ gravity to more general higher-curvature theories (e.g., Lanczos--Lovelock gravity) and to entropy functionals with explicit dependence on curvature invariants may clarify to what extent the `slope controls coupling' mechanism persists when the horizon entropy becomes curvature-sensitive or genuinely non-local.

In summary, the picture that emerges in this work is that horizon thermodynamics can consistently accommodate non-extensive (and, more generally, non-additive) system behavior while also placing stringent constraints on it. Once the local Clausius balance is taken seriously, the entropy slope becomes the operationally relevant quantity, while topology, where physically or mathematically admissible, provides a systematic means of calibrating the associated area scale. Taken together, these ingredients make generalized area-law entropies quantitatively constrained and observationally testable deformations of the Bekenstein--Hawking law.

\section*{Acknowledgments}
G L acknowledges the Istituto Nazionale di Alta Matematica (INdAM), Gruppo Nazionale di Fisica Matematica.
P J acknowledges support from the Czech Science Foundation Grant Agency (GA\v{C}R), Grant No. 25-18 105S.

\section*{Data availability statement}
All data that support the findings of this study are included within the article (and any supplementary files).

\section*{Author contributions}
\noindent\textbf{Marco Figliolia} 0009-0003-2313-8045\\
Conceptualization (equal), Formal analysis (equal), Methodology (equal), Project administration (equal), Visualization (equal), Writing -- original draft (equal), Writing -- review \& editing (equal)

\noindent\textbf{Petr Jizba}\\
Conceptualization (equal), Investigation (equal), Methodology (equal), Writing -- original draft (equal), Writing -- review \& editing (equal)

\noindent\textbf{Gaetano Lambiase} 0000-0001-7574-2330\\
Conceptualization (equal), Validation (equal), Writing -- review \& editing (equal)


\clearpage
\appendix
\renewcommand{\theequation}{\arabic{equation}}

\section*{Appendix. Gauss--Bonnet entropy as a topological counterterm}
\setcounter{equation}{97}

In this appendix we illustrate how the calibration logic discussed in section~\ref{sec:TCP} operates in a theory where the gravitational Lagrangian already contains a purely topological density. A standard example is Einstein gravity supplemented by the four-dimensional Gauss--Bonnet density
\begin{equation}
  \mathcal{L} \ = \ \frac{1}{16\pi G}\,\big(R\ + \ \alpha\,\mathcal G\big)\, ,
  \qquad
  \mathcal G\ = \ R_{abcd}R^{abcd} \ - \ 4R_{ab}R^{ab}\ + \ R^2\, ,
\end{equation}
where $\mathcal G$ is the quadratic Gauss--Bonnet term. In four dimensions its integral is proportional to the Euler characteristic, so the equations of motion remain those of Einstein gravity while the Wald entropy acquires an additive topological contribution~\cite{WaldEntropy,IyerWald1994}
\begin{equation}
  S_{\rm GB}\ = \ \frac{2\pi\,\alpha}{G}\,\chi(\Sigma)\, .
\end{equation}
In particular, $S_{\rm GB}$ provides exactly the kind of slope-free topological counterterm envisaged by the TCP: since $dS_{\rm GB}/dA=0$, it shifts the entropy zero point and its global topological dependence without altering the slope $s_0$ and hence the local Jacobson equations.

In this setting the TCP logic can be used in a slightly different way: the Gauss--Bonnet contribution calibrates the total entropy at a fiducial area without modifying the local field equations. Concretely, define
\begin{equation}
  S_{\rm tot}(A,\chi)
 \ = \  S(A) \ + \ S_{\rm GB}(\chi)
  \ = \  \eta\Big(\frac{A}{4G}\Big)^{\delta}
    \ + \ \frac{2\pi\alpha}{G}\,\chi\, ,
\end{equation}
and choose a reference spherical section with $\chi=2$ and area $A_{\rm ren}$, naturally tied to the intrinsic curvature scale by $A_{\rm ren}=A_0(2)=8\pi/\tilde R_*$, where $A_0(\chi)$ is the Gauss--Bonnet reference area introduced in equation~\eqref{eq:A0-chi-TCP}. Imposing slope matching to the physical Newton coupling,
\begin{equation}
  \frac{dS_{\rm tot}}{dA}\Big|_{A_{\rm ren}}
  \ = \ \frac{1}{4\,G_{\rm phys}}\, ,
\end{equation}
and area-law normalization on the sphere
\begin{equation}
  S_{\rm tot}(A_{\rm ren};\chi=2)
  \ = \ \frac{A_{\rm ren}}{4\,G_{\rm phys}}\, ,
\end{equation}
one finds
\begin{equation}
  \eta
 \ =  \ \frac{G}{\delta\,G_{\rm phys}}
    \left(\frac{A_{\rm ren}}{4G}\right)^{1-\delta}\, ,
  \qquad
  \alpha
  \ = \  \frac{G}{16\pi}\,\frac{A_{\rm ren}}{G_{\rm phys}}
    \left(1 \ - \ \frac{1}{\delta}\right)\, .
\end{equation}
Equivalently, for given $(\alpha,G_{\rm phys})$ the calibration fixes $\delta$ and hence $\eta$. This example shows explicitly how the TCP logic can also relate non-extensive entropy parameters to a topological Gauss--Bonnet coupling in theories whose local dynamics remain Einsteinian.



\bibliographystyle{iopart-num}
\bibliography{references}

\end{document}